\documentclass[twocolumn,apj,numberedappendix,appendixfloats]{openjournal}
\usepackage[utf8]{inputenc}

\usepackage{amsmath}
\usepackage{amssymb}
\usepackage{graphicx}
\usepackage{fontawesome}
\usepackage{bm}
\usepackage{natbib}
\usepackage{hyperref}
\hypersetup{
    colorlinks=true,
    linkcolor=blue,
    filecolor=blue,
    urlcolor=blue,
    citecolor=blue,
    }

\usepackage{booktabs}
\usepackage{colortbl}
\usepackage[dvipsnames]{xcolor}
\usepackage{multirow}
\usepackage{algorithm}
\usepackage[noend]{algpseudocode}

\begin{document}

\journalinfo{The Open Journal of Astrophysics}
\submitted{submitted XXX; accepted YYY}

\shorttitle{Differentiable \& Accelerated Spherical Wavelets}
\shortauthors{Price, Polanska, Whitney, McEwen}

\title{Differentiable and accelerated directional wavelet transforms \\ on the sphere and ball}

\author{M.~A.~Price $^{\star,1}$}
\author{A.~Polanska $^{1}$}
\author{J.~Whitney $^{1}$}
\author{J.~D.~McEwen $^{\dagger1,2}$}

\affiliation{$^1$ Mullard Space Science Laboratory (MSSL), University College London (UCL), Holmbury St Mary, Dorking, Surrey RH5 6NT, UK}
\affiliation{$^2$ Alan Turing Institute, Euston Road, London NW1 2DB, UK}
\thanks{$^\star$ E-mail: \nolinkurl{m.price.17@ucl.ac.uk} \\
  $\dagger$ E-mail: \nolinkurl{jason.mcewen@ucl.ac.uk}}

\begin{abstract}
  Directional wavelet dictionaries are hierarchical representations which efficiently capture and segment information across scale, location and orientation. Such representations demonstrate a particular affinity to physical signals, which often exhibit highly anisotropic, localised multiscale structure. Many physically important signals are observed over spherical domains, such as the celestial sky in cosmology. Leveraging recent advances in computational harmonic analysis, we design new highly distributable and automatically differentiable directional wavelet transforms on the 2-dimensional sphere $\mathbb{S}^2$ and 3-dimensional ball $\mathbb{B}^3=\mathbb{R}^+\times\mathbb{S}^2$ (the space formed by augmenting the sphere with the radial half-line). We observe up to a $300$-fold and $21800$-fold acceleration for signals on the sphere and ball, respectively, compared to existing software, whilst maintaining 64-bit machine precision. Not only do these algorithms dramatically accelerate existing spherical wavelet transforms, the gradient information afforded by automatic differentiation unlocks many data-driven analysis techniques previously not possible for these spaces.
  We publicly release both \texttt{S2WAV} {\href{https://github.com/astro-informatics/s2wav}{\faGithub}} and \texttt{S2BALL} {\href{https://github.com/astro-informatics/baller}{\faGithub}}, open-sourced \texttt{JAX} libraries for our transforms that are automatically differentiable and readily deployable both on and over clusters of hardware accelerators (\emph{e.g.} GPUs \& TPUs).
\end{abstract}


\maketitle

\section{Introduction}
Many fields of research fundamentally depend on the distillation of scientifically pertinent information from data that lives on spherical manifolds; that is data which lives on the 2-dimensional sphere $\mathbb{S}^2$. In many cases such data is radially distributed, and therefore lives on the 3-dimensional ball $\mathbb{B}^3=\mathbb{R}^+\times\mathbb{S}^2$ (the space formed by augmenting the sphere with the non-negative radial half-line). The diversity of such fields is remarkable, ranging from: quantum chemistry \citep{ritchie:1999,choi:1999}, to molecular modelling and protein prediction \citep{boomsma:2017:spherical,jumper:alphafold}, to biomedical imaging \citep{tuch:2004,goodwin:2022:dMRI}, to geophysics and planetary science \citep{audet:2010, simons:2011, marignier:2020:probability} to atmospheric and climate physics \citep{racah:2017:extremeweather, weyn:2020, ravuri:2021}, and to the wider cosmos \citep{mcewen:2008:ng,mcewen:2007:isw2, price:massmapping_uq_sphere, wallis:mass_mapping_sks,loureiro:2022:almanac}.
Increasingly often spherical data is encountered in modern computer vision tasks, \emph{e.g.} monocular depth estimation and semantic segmentation of $360^\circ$ images \citep{jiang:2019, zhang:2019:orientation, ocampo:2023:scalable}, or 3D object retrieval \citep{kondor:2018:clebsch, esteves:2020:spin,cobb:efficient_generalized_s2cnn}.

In recent years machine learning techniques have become increasingly widespread within these fields \citep[see \emph{e.g.}][]{ntampaka:2019, bronstein:2021:geometric,dawes:2023}. Despite numerous advantages afforded by machine learning approaches, their efficacy is typically predicated on an abundance of training data. In many fields such data is not available, and simulation of such data is simply not feasible; astrophysics being a classic example. When data is limited, existing tools can be reimagined and employed to create hybrid methods, at great effect. One such class of tools is that of the wavelet transform and its derivatives \citep[see \emph{e.g.}][]{bruna:2013:invariant, mcewen:scattering, pedersen:2023:learnable}.

Wavelet theory on the sphere is a mature field of study, with a plethora of associated wavelet representations, including spherical; wavelets \citep{schroder:1995, antoine:1998,antoine:1999}, needlets \citep{geller:2008, baldi:2009}, curvelets \citep{starck:2006, starck:2009,chan:s2let_curvelets}, and ridgelets \citep{mcewen:s2let_ridgelets} each tailored towards specific applications. Of particular interest are scale-discretised spherical wavelets \citep{leistedt:s2let_axisym,mcewen:s2let_spin}, which satisfy important quasi-exponential localisation and asymptotic uncorrelation properties \citep{mcewen:s2let_localisation}, and can be composed to form isometrically invariant representations which are stable to diffeomorphisms, \textit{i.e.} effective representations for learning on the sphere \citep{mcewen:scattering}. Slepian wavelet representations on the sphere have also been developed \citep{roddy:slepian_wavelets, roddy:2023:riemannian} which are particularly well suited to the masked sky, often encountered in astrophysics.

A subset of these wavelet representations have been lifted to the ball, to facilitate the analysis of spherically symmetric and radially distributed data. The exact properties of such wavelets is, in part, determined by the choice of radial discretisation. Radial needlets \citep{durastanti:2014} adopt exponential radial basis functions whereas isotropic wavelets \citep{lanusse:2012} adopt spherical Bessel functions, and are consequently built on the Fourier-Bessel transform -- which is encountered throughout cosmology \citep{abramo:2010, rassat:2012}. As the spherical Bessel function does not admit exact quadrature \citep{lemoine:1994:discrete}, these isotropic wavelets are not exact. Instead, Fourier-Laguerre wavelets \citep{leistedt:flaglets} select Laguerre polynomials with which to tile the radial half-line, and leverage Gauss-Laguerre quadrature to provide theoretically exact transforms. Interestingly, the spherical Bessel transform may be evaluated exactly by first projecting onto Laguerre polynomials, for signals bandlimited in the Fourier-Laguerre sense \citep{leistedt:flaglets}. Slepian wavelets have also been extended to the ball \citep{khalid:ball_slepian} which are built upon a Fourier-Laguerre spectral decomposition.

Classically, wavelets representations have been effectively applied as compressive sensing regularisers when solving inverse problems \citep[\emph{e.g.}][]{carrillo:purify_icassp, pratley:purify_real_data, price:massmapping_uq_hypothesis}. Wavelets may be composed to form expressive scattering representations, which are both effective summary statistics and statistical generative models for highly non-Gaussian textures \citep{mallat:2012, mallat:2020:phase, allys:2019:rwst,allys:2020:new,cheng:2020:new,zhang:2021:maximum, eickenberg:2022:wavelet, mcewen:scattering, price:2023:strings}. Perhaps less well known are their applications for data compression \citep{balan:2009:painless, mcewen:szip} and the synthesis of multifractal fields \citep{robitaille:2020}. Recently, researchers have achieved state-of-the-art performance by embedding wavelet filters directly within machine learning models \citep{huang:2017, liu:2019}, multiscale conditioning of diffusion-based generative models \citep{guth:2022}, and by solving partial differential equations with wavelet neural operators \citep{gupta:2021, tripura:2022}.

Many of these technological advances are fundamentally reliant on readily accessible gradient information, which is necessary for back-propagation during model training. A further advantage of many modern technologies is their affinity for high throughput evaluation; hence, they benefit greatly from deployment on hardware accelerators (\emph{e.g.} GPUs and TPUs). Existing software packages which provide wavelet transforms on the sphere, \texttt{S2LET} {\href{https://github.com/astro-informatics/s2let}{\faGithub}}, and ball, \texttt{FLAGLET} {\href{https://github.com/astro-informatics/flaglet}{\faGithub}}, are not engineered with this in mind, and provide neither differentiability nor acceleration. Consequently, it has not been possible to integrate these wavelet transforms with modern machine learning techniques on the sphere. To harness the potential of wavelets for next-generation spherical data-analysis techniques, new algorithms and software are needed.

In this work we design spherical wavelet transforms which overcome these fundamental limitations. Building upon recently released GPU accelerated spherical harmonic and Wigner transforms, provided by the \texttt{S2FFT} {\href{https://github.com/astro-informatics/s2fft}{\faGithub}} package \citep{price:s2fft}, we develop highly parallel algorithms for the scale-discretised wavelet transform on the sphere and ball. We implement these spherical wavelet transforms in \texttt{JAX}, a differentiable \texttt{Python} library developed by Google for high-performance deep learning research \citep{jax:2018:github}. Our transforms are engineered so as to provide efficient automatic differentiation and be highly distributable both on and over hardware accelerators; hence, facilitating the integration the future integration of wavelet techniques with modern machine learning technologies over spherically symmetric spaces. To maximise accessibility our algorithms are designed to be sampling agnostic, providing support for McEwen-Wiaux \citep{mcewen:fssht}, Driscoll-Healy \citep{driscoll:1994}, and HEALPix \citep{gorski:2005} sampling of the sphere at launch. Differentiability notwithstanding, the acceleration afforded by our algorithms alone facilitates previously infeasible analysis techniques, \emph{e.g.} sampling methods and machine learning, with myriad applications from molecular modelling to the study of the cosmos.

The remainder of this article is structured as follows. In Section \ref{sec:background} we review mathematical background for harmonic analysis, with wavelet analysis provided in Section \ref{sec:multiscale_analysis}. In Section \ref{sec:s2wav} we develop the directional wavelet transform on the sphere and outline our associated software package \texttt{S2WAV} {\href{https://github.com/astro-informatics/s2wav}{\faGithub}}. Subsequently, in Section \ref{sec:baller} we abstract to the ball and outline our associated software package \texttt{S2BALL} {\href{https://github.com/astro-informatics/baller}{\faGithub}}. Finally, in Section \ref{sec:conclusions} we draw conclusions and make closing remarks.

\section{Harmonic analysis on the sphere, rotation group, and ball} \label{sec:background}
In this section we review harmonic analysis on the two-sphere $\mathbb{S}^2$, rotation group SO(3), three-dimensional ball $\mathbb{B}^3 = \mathbb{R}^+ \times \mathbb{S}^2$, and rotational ball $\mathbb{H}^4 = \mathbb{R}^+ \times \text{SO(3)}$. We provide a very brief summary of necessary theory relating to spherical harmonics, Wigner functions, and Fourier-Laguerre polynomials. Our discussion covers both axisymmetric and directional convolutions on the aforementioned manifolds, for functions of arbitrary spin. We constrain this discussion to continuous transforms so as to remain sampling agnostic. In later sections we connect to corresponding transforms over discrete spaces, bridging the gap to practical applications.

\subsection{Functions on the sphere}
Spin-$s$ functions on the sphere ${}_sf\in\text{L}^2[\mathbb{S}^2]$ are characterised by an additional $\mathbb{U}(1)$ symmetry such that they transform by \citep{newman:1966,goldberg:1967}
\begin{equation} \label{eq:u1sym}
  {}_sf^{\prime} = e^{-is\chi} {}_sf,
\end{equation}
under right-handed rotations $\chi \in [0, 2\pi)$ in the tangent plane centered at $\omega=(\vartheta,\varphi)$, for longitude $\varphi \in [0, 2\pi)$ and colatitude $\vartheta \in [0,\pi]$. The spin-weighted spherical harmonics ${}_sY_{\ell m}(\omega) : \mathbb{S}^2 \rightarrow \mathbb{C}$ form the canonical orthogonal basis for square integrable functions $\text{L}^2[\mathbb{S}^2]$ on the sphere for natural $\ell\in\mathbb{N}$ and integers $m,s \in \mathbb{Z}$ such that $|m|,|s| \leq \ell$. Any spin-$s$ function ${}_sf \in \text{L}^2[\mathbb{S}^2]$ may be decomposed into this representation such that
\begin{equation} \label{eq:fswsh}
  {}_s\hat{f}
  = \langle \, {}_sf, \,{}_sY_{\ell m} \, \rangle
  = \int_{\mathbb{S}^2} \text{d}\Omega(\omega) \: {}_sf(\omega) \: {}_sY_{\ell m}^* (\omega),
\end{equation}
where $\Omega(\omega) = \sin\vartheta\,\text{d}\vartheta \,\text{d}\varphi$ is the standard invariant measure on the sphere. By orthogonality and completeness of the spin-weighted spherical harmonics, ${}_sf$ may be reconstructed exactly by
\begin{equation} \label{eq:iswsh}
  {}_sf
  = \sum_{\ell \, \in \, \mathbb{N}} \, \sum_{|\,m\,| \, \leq \, \ell} \: {}_s\hat{f}_{\ell m} \: {}_sY_{\ell m}.
\end{equation}

For practical calculation this infinite summation is truncated at a maximum harmonic degree $L$ such that ${}_s\hat{f}_{\ell m} = 0$ for all $\ell \geq L$. The terminology is then such that we say the function ${}_sf$ is bandlimited at $L$. Efficient GPU algorithms have been developed to perform discretised versions of the transforms given by Equations \ref{eq:fswsh} and \ref{eq:iswsh} \citep[see \texttt{S2FFT} {\href{https://github.com/astro-informatics/s2fft}{\faGithub}},][]{price:s2fft}.

\subsection{Functions on the rotation group}
As they will play a pivotal role in the construction of directional wavelet transforms on the sphere we introduce functions on the rotation group, again very briefly. The Wigner $D$-functions $D^{\ell}_{mn}(\rho) : \text{SO(3)} \rightarrow \mathbb{C}$ for natural $\ell \in \mathbb{N}$, integers $m,n \in \mathbb{Z}$ such that $|\,m\,|,\,|\,n\,|\,\leq \ell$, and $zyz$ Euler angles $\rho = (\alpha, \beta, \gamma)$ form an irreducible unitary representation of the rotation group SO(3) in three dimensions \citep{varshalovich:1989}.

Any square integrable function $f\in\text{L}^2[\text{SO(3)}]$ on the rotation group may, as before, be decomposed into this representation such that
\begin{equation} \label{eq:forward_wigner_transform}
  \hat{f}
  = \langle \, f, \, D^{\ell}_{mn} \, \rangle
  = \int_{\text{SO(3)}} \text{d}\Omega(\rho) \: f(\rho) \: D^{\ell}_{mn}(\rho),
\end{equation}
where $\Omega(\rho) = \sin\beta\,\text{d}\alpha \, \text{d}\beta \,\text{d}\gamma$ is the usual invariant measure on the rotation group. By the orthogonality and completeness of the Wigner $D$-functions, $f$ may be reconstructed exactly by
\begin{equation} \label{eq:inverse_wigner_transform}
  f = \sum_{\ell \, \in \, \mathbb{N}} \: \frac{2\ell+1}{8\pi^2} \:
  \sum_{|\,m\,| \,\leq \,\ell} \, \sum_{|\,n\,| \,\leq \,\ell} \:
  \hat{f}^{\ell}_{mn} \: D^{\ell*}_{mn}.
\end{equation}

As for the spin-weighted spherical harmonic transforms, the infinite summation is bandlimited at $L$. Efficient GPU algorithms have been developed to perform discretised version of the Wigner transform, which we leverage \citep[see \texttt{S2FFT} {\href{https://github.com/astro-informatics/s2fft}{\faGithub}},][]{price:s2fft}.

\subsection{Functions on the ball} \label{sec:funcs_on_ball}
To extend harmonic analysis radially one must introduce basis functions along the positive half-line $\mathbb{R}^+$. As the canonical choice of atlas on spherical spaces are the spherical polar co-ordinates, which are separable into angular and radial components, we are free to straightforwardly adopt either spin-weighted spherical harmonics or Wigner $D$-functions for the angular components and independent basis functions along the radial half-line.

The canonical choice of radial basis functions are the Bessel functions which, when combined with the angular basis functions, produce the spherical Bessel functions. Unfortunately, these functions do not afford exact transforms and can be numerically unstable in practical settings. Fortunately, alternate basis functions with desirable qualities are available. The Laguerre basis functions $K_p(r): \mathbb{R}^+ \rightarrow \mathbb{R}^+$, which are orthogonal by Gram-Schmidt and are straightforwardly complete,  are defined to be
\begin{equation}
  K_p(r)
  = \sqrt{\frac{p!}{(p+2)!}} \: \frac{e^{\frac{-r}{2\tau}}}{\sqrt{\tau^3}} \: L_p^{(2)}\Big(\frac{r}{\tau}\Big),
\end{equation}
where $L_p^{(2)}$ are the $p^{\text{th}}$-associated $2^{\text{nd}}$-order Laguerre polynomials, for natural $p\in\mathbb{N}$, and where $\tau \in \mathbb{R}^+$ is a scaling parameter. Any square integrable function ${}_sf\in\text{L}^2[\mathbb{R}^+]$ may be decomposed into this representation such that
\begin{equation} \label{eq:flag}
  {}_s\hat{f}
  = \langle \, {}_sf, \, K_p \, \rangle
  = \int_{\mathbb{R}^+} \text{d}rr^2 \: {}_sf(r)\:K_p(r),
\end{equation}
and for the sampling theorem presented by \citet{leistedt:flaglets} may be recovered exactly by
\begin{equation} \label{eq:ilag}
  {}_sf(r)
  = \sum_{p \, \in \, \mathbb{N}} {}_s\hat{f}_p \: K_p(r).
\end{equation}

Most real-world functions are, to a good approximation, radially bandlimited which is to say that ${}_s\hat{f}_p = 0$ for all $p > P$. This is the same as bandlimiting in the angular components, where instead we say a function $f$ is radially bandlimited at $P$. Adopting Gauss-Laguerre quadrature to compute the integral of Equation \ref{eq:flag} discrete Laguerre transforms can be evaluated exactly \citep{leistedt:flaglets,mcewen:flaglets_sampta}.

One can then straightforwardly compose radial and angular basis functions to form a set of basis functions on the ball. Suppose for the angular components we adopt the spin-weighted spherical harmonics, then our overall basis functions are defined to be
\begin{equation}
  {}_sZ_{\ell mp}(b) = K_p(r) \: {}_sY_{\ell m}(\omega),
\end{equation}
for $b = (r,\omega) \in \mathbb{B}^3 = \mathbb{R}^+ \times \mathbb{S}^2$. These basis functions inherit the characteristics of their constituents, and are therefore trivially orthogonal and complete. Consequently, any square integrable spin-$s$ function ${}_sf \in \text{L}^2[\mathbb{B}^3]$ may be decomposed into their spherical-Laguerre representation such that
\begin{equation} \label{eq:forward_spherical_laguerre}
  {}_s\hat{f}
  = \langle \, {}_sf, \, {}_sZ \, \rangle
  = \int_{\mathbb{B}^3} \text{d}\Omega(b) \: {}_sf(b) \: {}_sZ^*_{\ell mp}(b),
\end{equation}
where $\text{d}\Omega(b) = \text{d}\Omega(r) \text{d}\Omega(\omega) = r^2 \, \sin\vartheta \, \text{d}r \, \text{d}\vartheta \, \text{d}\varphi$ is the standard invariant measure on the ball. It then follows that ${}_sf$ may be reconstructed exactly by
\begin{equation} \label{eq:inverse_spherical_laguerre}
  {}_sf
  = \sum_{p,\ell\,\in\,\mathbb{N}} \, \sum_{|\,m\,|\,\leq \, \ell} \:
  {}_s\hat{f}_{\ell mp} \: {}_sZ_{\ell mp},
\end{equation}
where in practice infinite summations over $\ell,p$ are angularly and radially bandlimited by $L,P$ respectively. Building on previous work \citep{leistedt:flaglets,mcewen:flaglets_sampta,leistedt:flaglets_spin}, in this work we develop differentiable and highly accelerated GPU algorithms to evaluate this spherical-Laguerre transform, which we extend to the directional setting.

\subsection{Directional convolutions on the sphere}
Consider a rotation operator $\mathcal{R}_{\rho}$ with action
\begin{equation}
  (\mathcal{R}_{\rho} \: {}_sf)(\omega)
  = e^{-is\vartheta} \: {}_sf(\,\mathcal{R}_{\rho}^{-1}\omega),
\end{equation}
when applied to a spin-$s$ square integrable function on the sphere. The exponential term here comes from the additional $\mathbb{U}(1)$ symmetry discussed in Equation \ref{eq:u1sym} \citep{mcewen:s2let_spin}. By noting the additive property of the Wigner $D$-functions, a rotated function permits a harmonic representation
\begin{equation}
  (\mathcal{R}_{\rho} \: {}_sf)_{\ell m} = \sum_{|\,n\,|\,\leq\,\ell} \: D^{\ell}_{mn} \: {}_sf_{\ell n}.
\end{equation}

Suppose one would like to convolve two spin-$s$ square integrable functions ${}_sf,{}_sg \in \text{L}^2[\mathbb{S}^2]$. Conceptually, this is given as the product between ${}_sf$ and $\mathcal{R}_{\rho}\,{}_sg$ for all possible Euler angles $\rho \in \text{SO(3)}$. Mathematically, this reads
\begin{align} \label{eq:direct_conv}
  ({}_sf \circledast {}_sg)(\rho)
   & = \langle \, {}_sf, \, \mathcal{R}_{\rho}\,{}_sg \,\rangle \nonumber                                                \\
   & = \int_{\mathbb{S}^2} \text{d}\Omega(\omega) \: {}_sf(\omega) \: \big ( \mathcal{R}_{\rho}\,{}_sg \big )^*(\omega),
\end{align}
where $\circledast$ is the directional convolution. As this is a function on the rotation group, we may consider the Wigner representation for simplicity
\begin{equation} \label{eq:angular_conv}
  ({}_sf \circledast {}_sg)^{\ell}_{mn} = \frac{8\pi^2}{2\ell+1}\: {}_sf_{\ell m} \: {}_sg_{\ell n}^*,
\end{equation}
from which the convolved function can be recovered exactly by \citep{mcewen:2007:fast,mcewen:s2let_spin, cobb:efficient_generalized_s2cnn}
\begin{equation}
  ({}_sf \circledast {}_sg)(\rho)
  =
  \sum_{\ell \, \in \, \mathbb{N}} \, \sum_{|\,m,n\,|\,\leq\,\ell} \:
  {}_sf_{\ell m} \: {}_sg_{\ell n}^* \: D^{\ell*}_{mn}.
\end{equation}

When we restrict the possible rotations to the sphere, that is when $\rho = (\alpha, \beta, \gamma) \rightarrow (\alpha, \beta, 0) = (\vartheta, \varphi) = \omega$, and when ${}_s g$ is axisymmetric ${}_s g_{\ell m}={}_sg_{\ell 0}\delta_{m0}$, this convolution becomes the axisymmetric convolution on the sphere $\mathbb{S}^2$.

\subsection{Convolutions on the radial half-line}
Further consider a generalised radial translation\footnote{From our construction of the spherical-Laguerre basis functions this is closer to a translation across Laguerre polynomials, however we will refer to this as radial translation throughout this article.} $\mathcal{T}_r$ for $r\in\mathbb{R}^+$ defined by its application to the spherical-Laguerre basis functions
\begin{equation}
  ( \mathcal{T}_{r^\prime} \: K_p)(r)
  = K_p(r^{\prime})\: K_p(r),
\end{equation}
presented in \citet{mcewen:flaglets_sampta} with action on square integrable functions $f\in\mathbb{R}^+$ given by
\begin{align}
  (\mathcal{T}_{r^{\prime}} \: f)(r)
   & = \sum_{\,p \,\in \,\mathbb{N}\,} \: \hat{f}_p \: K_p(r^{\prime}) \: K_p(r) \nonumber \\
  \Rightarrow (\,\mathcal{T}_{r^{\prime}} \: f\,)_p
   & = K_p(r^{\prime}) \: \hat{f}_p.
\end{align}
Using this definition of the translation operator we can define the convolution between two functions $f,g\in\text{L}^2[\mathbb{R}^+]$ to be given as
\begin{align} \label{eq:radial_conv}
  (f \: \star \: g)(r)
   & = \langle \, f, \, \mathcal{T}_r \, g\rangle \nonumber                         \\
   & = \int_{\mathbb{R}^+} \text{d}rr^2 \: f(r) \: (\mathcal{T}_{r^\prime} \,g)(r),
\end{align}
which straightforwardly leads to
\begin{equation}
  (f \: \star \: g)_p = f_p \: g_p.
\end{equation}

\subsection{Directional convolutions on the ball}
Finally, consider the combined 4-dimensional symmetry transformation $\mathcal{L}_h = \mathcal{T}_r\mathcal{R}_\rho$ for $h = (r, \rho) \in \mathbb{H}^4 = \mathbb{R}^+ \times \text{SO(3)}$, which describes the complete space of rotations and translations present in our system. The total convolution of two spin-$s$ functions $f,g \in \text{L}^2[\mathbb{B}^3]$ is defined analogously to before by \citep{price:b3inv}
\begin{align}
  ({}_sf \: \circledast \: {}_sg)(h)
   & = \langle \, {}_sf, \, \mathcal{L}_h \, {}_sg\rangle \nonumber                        \\
   & = \int_{\mathbb{B}^3} \text{d}\Omega(b) \: {}_sf(b) \: (\mathcal{L}_h \, {}_sg)^*(b),
\end{align}
which is the product between $f$ and $g$ over the composite space of 4-dimensional translations and rotations. Comparing against Equations \ref{eq:angular_conv} and \ref{eq:radial_conv} one finds
\begin{equation} \label{eq:wig_lag_conv_rep}
  ({}_sf \: \circledast \: {}_sg)^{\ell}_{mnp} = \frac{8\pi^2}{2\ell+1}\:{}_sf^{\ell}_{mp} \: {}_sg^{\ell*}_{np},
\end{equation}
from which the convolved function can be recovered in pixel-space by
\begin{equation}
  ({}_sf \: \circledast \: {}_sg)(h) =
  \sum_{p, \ell\,\in\,\mathbb{N}} \,
  \sum_{|\,m,n\,|\,\leq\,\ell}\,
  \frac{8\pi^2}{2\ell+1}\:{}_sf^{\ell}_{mp} \: {}_sg^{\ell*}_{np} \:
  Q^{\ell*}_{mnp}(h),
\end{equation}
where we have overloaded $\circledast$ to denote the convolution under the general symmetry transformation $\mathcal{L}_{h}$.

Here we have implicitly defined the Wigner-Laguerre basis functions $Q^{\ell}_{mnp}$ which are orthogonal and complete on $\mathbb{H}^4$. By construction the Wigner-Laguerre functions are straightforwardly separable, hence the decomposition of square integrable functions $f \in \text{L}^2[\mathbb{H}^4]$  is given by
\begin{equation}
  \hat{f}
  = \langle \,f, \,Q^{\ell}_{mnp} \,\rangle
  = \int_{\mathbb{H}^4} \text{d}\Omega(h) \: f(h) \: Q^{\ell*}_{mnp}(h),
\end{equation}
where $\Omega(h) = r^2 \sin\beta \, \text{d}r \, \text{d}\alpha \, \text{d}\beta \, \text{d}\gamma$ is the invariant measure on $\mathbb{H}^4$. As in previous settings, the original function may be exactly synthesised by
\begin{equation} \label{eq:wiglag_inverse}
  f
  = \sum_{p, \ell\,\in\,\mathbb{N}} \: \frac{2\ell+1}{8\pi^2} \:
  \sum_{|\,m,n\,|\,\leq\,\ell} \,
  \hat{f}^{\ell}_{mnp} \: Q^{\ell*}_{mnp},
\end{equation}
which is practically extremely expensive to evaluate. Previously, \citet{leistedt:flaglets,leistedt:flaglets_spin} developed algorithms to perform this transform. In this work, we redesign these algorithms to leverage recent advances in GPU accelerated and differentiable harmonic analysis \citep[see \texttt{S2FFT} {\href{https://github.com/astro-informatics/s2fft}{\faGithub}},][]{price:s2fft}.

\section{Wavelet transforms on the sphere and ball} \label{sec:multiscale_analysis}
In this section we discuss the wavelet analysis of signals on the sphere $\mathbb{S}^2$ and ball $\mathbb{B}^3 = \mathbb{R}^+ \times \mathbb{S}^2$. We formally define the directional wavelet transform on both the sphere and ball, leveraging much of the mathematics provided in Section \ref{sec:background}. Again we remain in the continuous setting.

\begin{figure*}
  \centering
  \includegraphics[width=0.19\textwidth,trim={5cm 3.2cm 5cm 3cm}, clip]{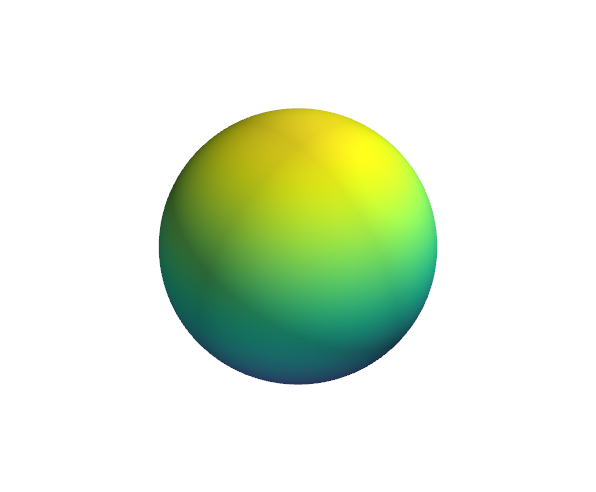}
  \includegraphics[width=0.19\textwidth,trim={5cm 3.2cm 5cm 3cm}, clip]{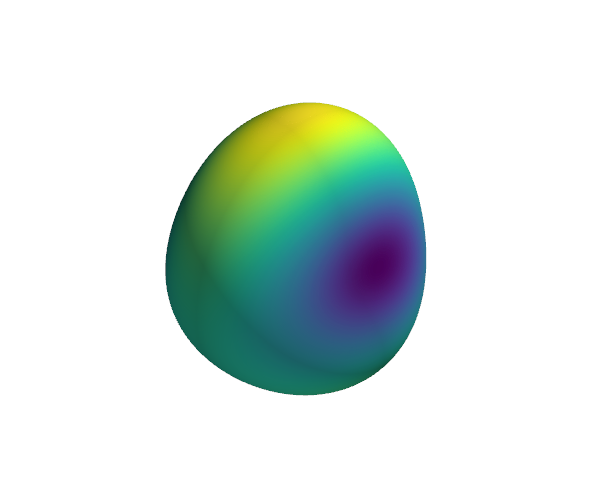}
  \includegraphics[width=0.19\textwidth,trim={5cm 3.2cm 5cm 3cm}, clip]{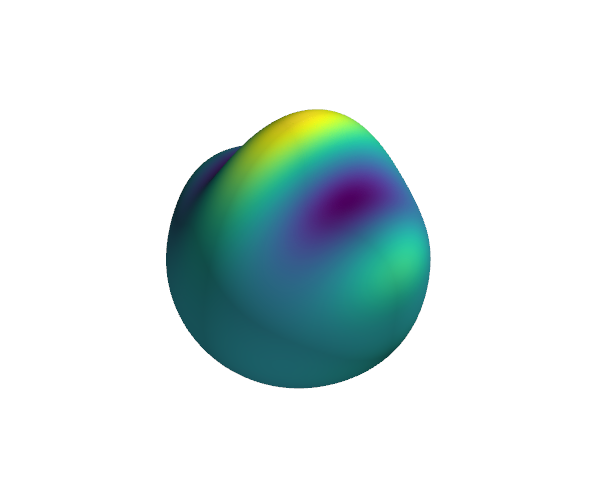}
  \includegraphics[width=0.19\textwidth,trim={5cm 3.2cm 5cm 3cm}, clip]{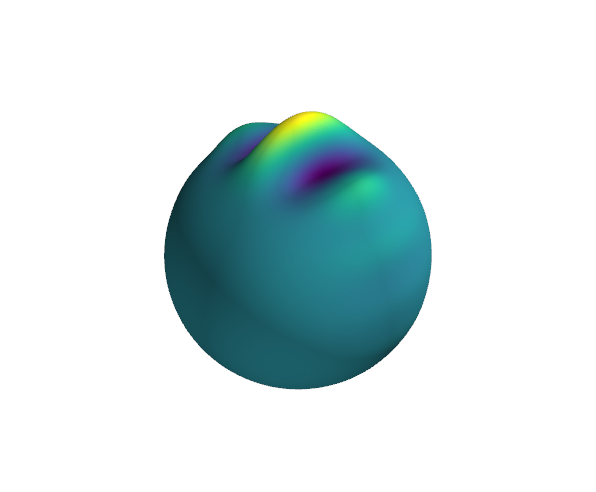}
  \includegraphics[width=0.19\textwidth,trim={5cm 3.2cm 5cm 3cm}, clip]{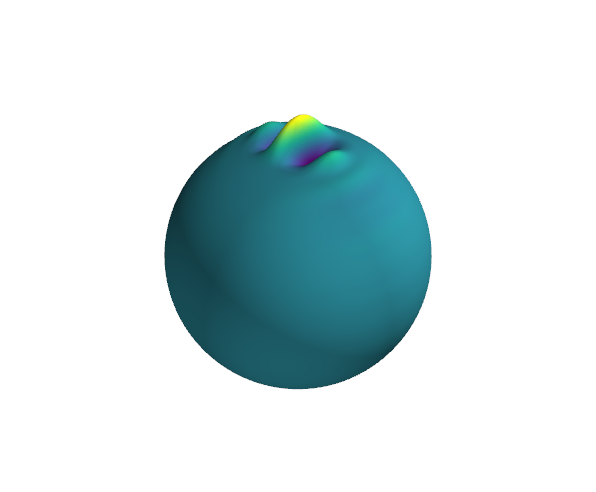}
  \put(-412,80){\large $\Psi^{1}$}
  \put(-315,80){\large $\Psi^{2}$}
  \put(-220,80){\large $\Psi^{3}$}
  \put(-120,80){\large $\Psi^{4}$}
  \put(-20,80){\large $\Psi^{5}$}
  \caption{Directional ($L=256, N=5$) scale-discretised wavelet filters on the sphere, for wavelet scales $j\in\lbrace1,\dots,5\rbrace$ from left to right. As the wavelet scale increases the filter becomes more highly localised, both in pixel and frequency space \citep{mcewen:s2let_localisation}. Notice the clearly elongated structure which gives rise to the desired directional sensitivity \citep{mcewen:s2let_spin}. With an azimuthal bandlimit of $N=5$ this filter bank is augmented by $9$ rotated filters, which are not shown here.} \label{fig:directional_wavelet_filters}
\end{figure*}

\subsection{Directional wavelet transform on the sphere} \label{sec:s2_wav_cont}
Spherical wavelet filters are square integrable bandlimited functions ${}_s\Psi^{(j)} \in \text{L}^2[\mathbb{S}^2]$ constructed to exhibit strong localisation properties in both the harmonic and spatial domain \citep[][]{mcewen:s2let_localisation}, as illustrated in Figure \ref{fig:directional_wavelet_filters}. Note that here $j \in \mathbb{N} \leq J$ denotes wavelet scale, which determines the harmonic degrees over which the wavelet has support (see Appendix \ref{sec:wav_construct}). The wavelet functions form a dictionary into which spin-$s$ functions ${}_sf\in\text{L}^2[\mathbb{S}^2]$ may be decomposed through the directional convolution
\begin{equation}
  W^{{}_s\Psi^j}(\rho)
  = ({}_sf \circledast {}_s\Psi^j)(\rho)
  = \langle {}_sf, \mathcal{R}_{\rho} \, {}_s\Psi^j \rangle,
\end{equation}
which from Equation \ref{eq:direct_conv} can be given in Wigner space by the expression
\begin{equation} \label{eq:wigner_wav_coeffs}
  (W^{{}_s\Psi^j})^{\ell}_{mn}
  = \frac{8\pi^2}{2\ell+1} \:
  {}_s\hat{f}_{\ell m} \:
  {}_s\Psi^{j*}_{\ell n}.
\end{equation}
To capture low frequency information an axisymmetric scaling function ${}_s\Phi \in \text{L}^2[\mathbb{S}^2]$ is introduced into which $f$ may be decomposed as
\begin{equation}
  W^{{}_s\Phi}(\omega)
  = ({}_sf \odot {}_s\Phi)(\omega)
  = \langle {}_sf, \mathcal{R}_{\omega} \, {}_s\Phi \rangle,
\end{equation}
where $\odot$ denotes the axisymmetric convolution. Again, this can be given in harmonic space by the expression
\begin{equation} \label{eq:harmonic_scal_coeffs}
  (W^{{}_s\Phi})_{\ell m} = \sqrt{\frac{4\pi}{2\ell +1}} {}_sf_{\ell m} \: {}_s\Phi^*_{\ell 0}.
\end{equation}
Provided a wavelet dictionary which satisfies the admissibility condition
\begin{equation}
  \frac{4\pi}{2\ell+1}| {}_s\Phi_{\ell 0}|^2
  + \frac{8\pi^2}{2\ell+1}
  \sum_{j=0}^J \sum_{|\,m\,| \, \leq \, \ell}
  | {}_s\Psi^j_{\ell m}|^2 = 1 \quad \forall \, \ell,
\end{equation}
of which there are many \citep[see \emph{e.g.}][]{leistedt:s2let_axisym,chan:s2let_curvelets,mcewen:s2let_ridgelets},
one may exactly reconstruct ${}_sf$ by
\begin{align}
  {}_sf
   & = \int_{\mathbb{S}^2} \text{d}\Omega(\omega) W^{{}_s\Phi}(\omega)(\mathcal{R}_{\omega}\Phi)(\omega^\prime) \nonumber                                         \\
   & + \sum_{j=0}^J \int_{\text{SO(3)}} \text{d}\Omega(\rho) W^{{}_s\Psi^j}(\rho)(\mathcal{R}_\rho\, W^{{}_s\Psi^j})(\omega^\prime). \label{eq:s2_cont_synthesis}
\end{align}
In general this transform can be expensive to compute, motivating the development of efficient algorithms. In this article we adopt scale-discretised wavelets \citep{wiaux:2007:sdw, leistedt:s2let_axisym, mcewen:s2let_spin}. These wavelets exhibit good harmonic and spatial localisation \citep{mcewen:s2let_localisation}, and permit exact synthesis, at least in such a case that a sampling theorem on the sphere and ball is provided \citep[see \emph{e.g.}][]{driscoll:1994, mcewen:fssht, mcewen:flaglets_sampta}.

\begin{figure*}
  \centering
  \includegraphics[width=0.19\textwidth,trim={5cm 3.2cm 5cm 3cm}, clip]{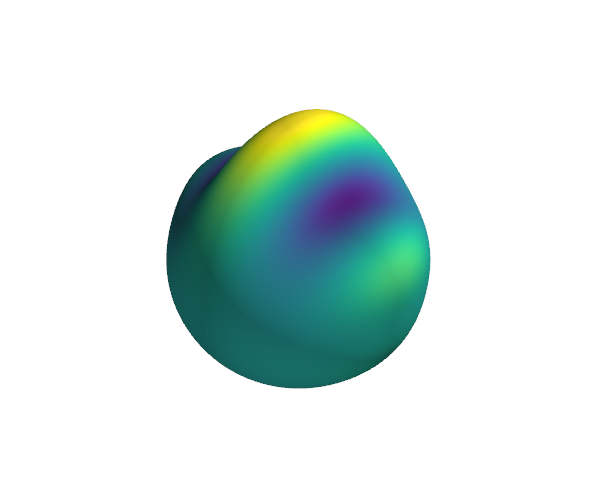}
  \includegraphics[width=0.19\textwidth,trim={5cm 3.2cm 5cm 3cm}, clip]{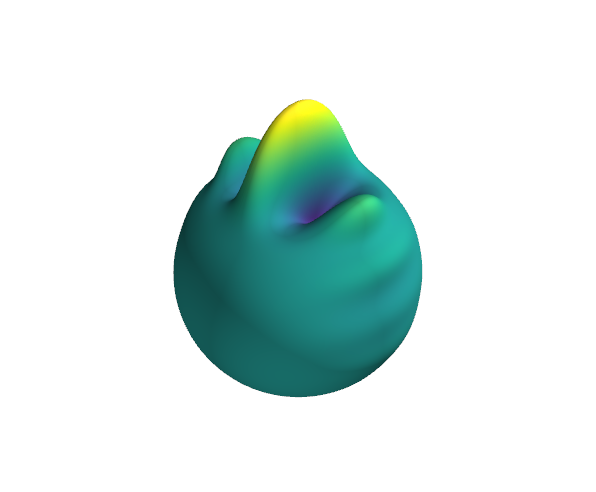}
  \includegraphics[width=0.19\textwidth,trim={5cm 3.2cm 5cm 3cm}, clip]{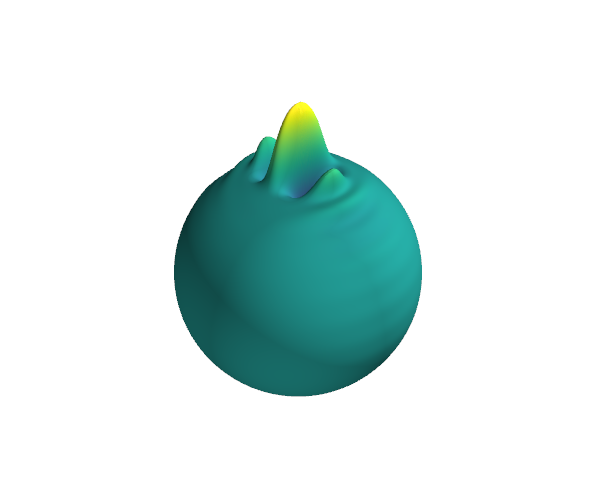}
  \includegraphics[width=0.19\textwidth,trim={5cm 3.2cm 5cm 3cm}, clip]{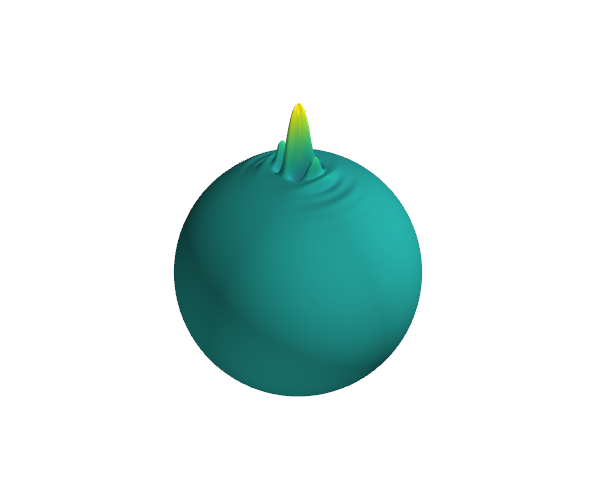}
  \includegraphics[width=0.19\textwidth,trim={5cm 3.2cm 5cm 3cm}, clip]{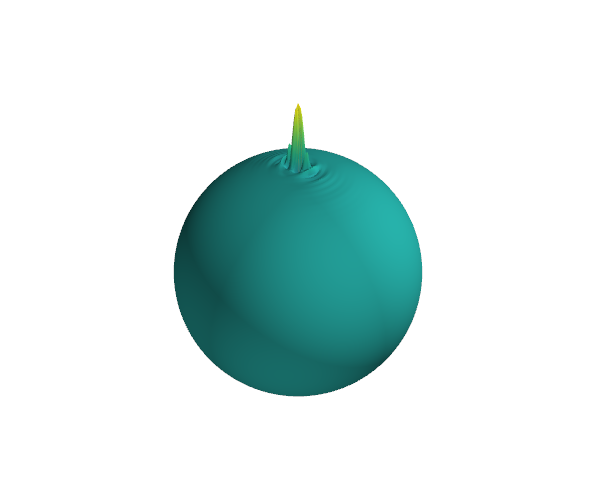}
  \put(-412,80){\large $\Psi^{3,3}$}
  \put(-315,80){\large $\Psi^{4,3}$}
  \put(-220,80){\large $\Psi^{5,3}$}
  \put(-120,80){\large $\Psi^{6,3}$}
  \put(-20,80){\large $\Psi^{7,3}$}
  \put(-505,36){\rotatebox{90}{\large $r=1$}}

  \includegraphics[width=0.19\textwidth,trim={5cm 3.2cm 5cm 3cm}, clip]{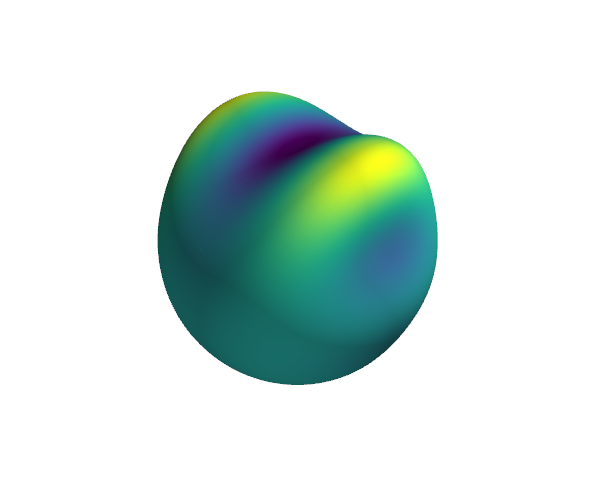}
  \includegraphics[width=0.19\textwidth,trim={5cm 3.2cm 5cm 3cm}, clip]{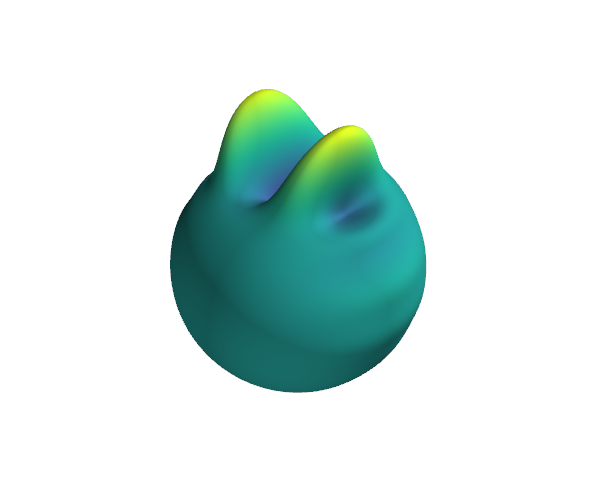}
  \includegraphics[width=0.19\textwidth,trim={5cm 3.2cm 5cm 3cm}, clip]{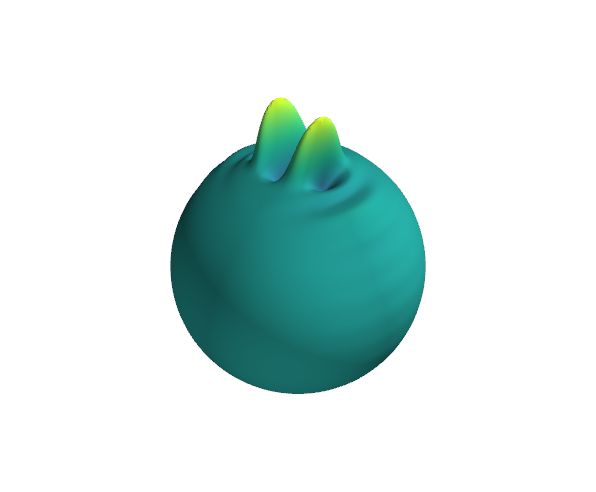}
  \includegraphics[width=0.19\textwidth,trim={5cm 3.2cm 5cm 3cm}, clip]{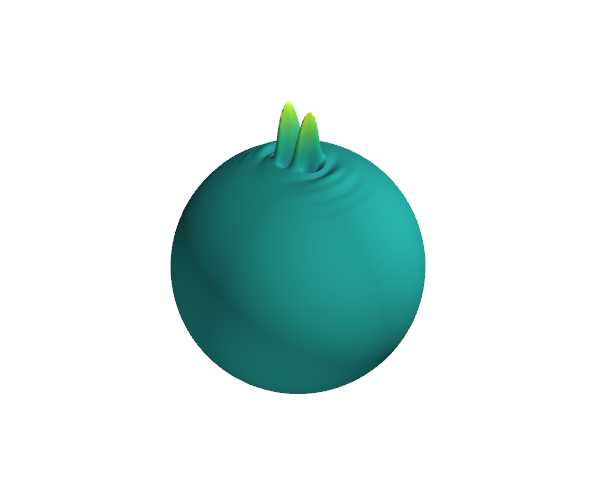}
  \includegraphics[width=0.19\textwidth,trim={5cm 3.2cm 5cm 3cm}, clip]{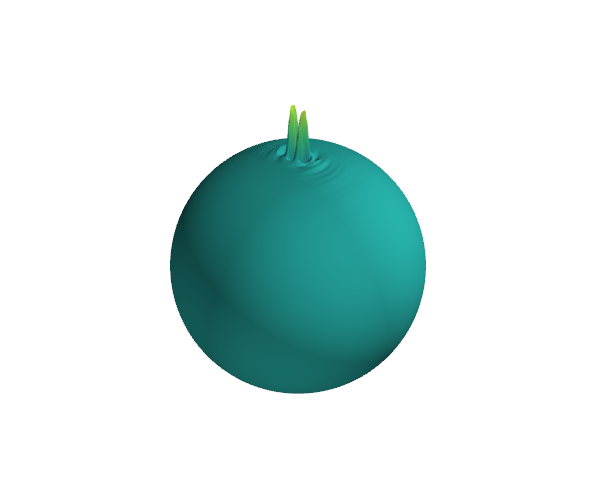}
  \put(-505,36){\rotatebox{90}{\large $r=2$}}

  \includegraphics[width=0.19\textwidth,trim={5cm 3.2cm 5cm 3cm}, clip]{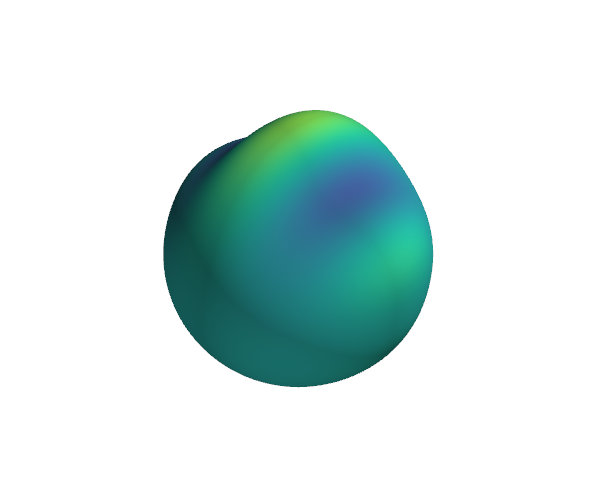}
  \includegraphics[width=0.19\textwidth,trim={5cm 3.2cm 5cm 3cm}, clip]{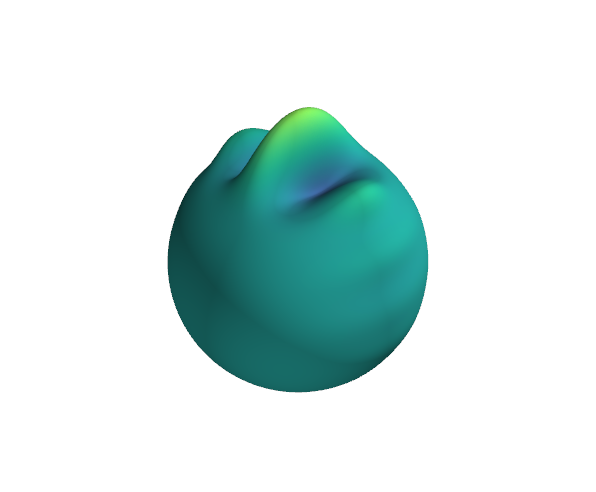}
  \includegraphics[width=0.19\textwidth,trim={5cm 3.2cm 5cm 3cm}, clip]{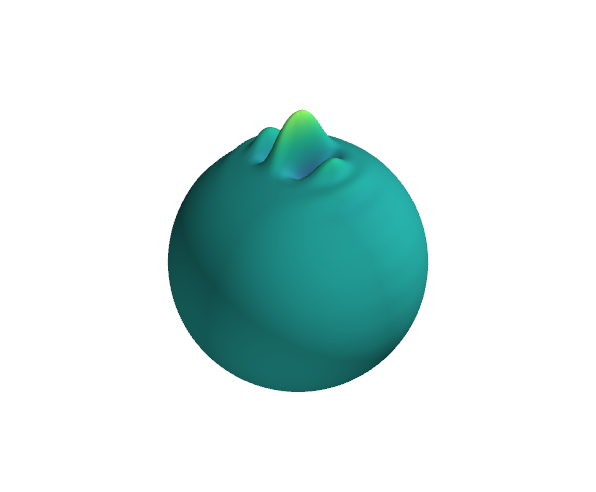}
  \includegraphics[width=0.19\textwidth,trim={5cm 3.2cm 5cm 3cm}, clip]{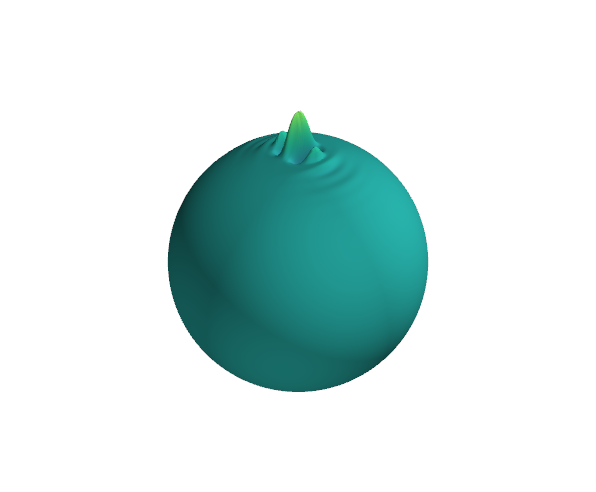}
  \includegraphics[width=0.19\textwidth,trim={5cm 3.2cm 5cm 3cm}, clip]{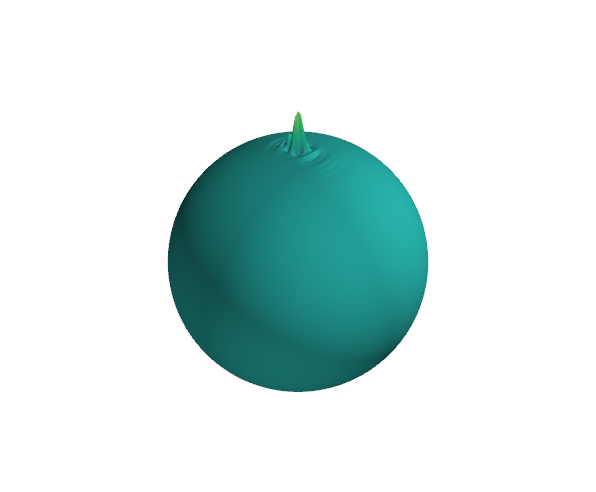}
  \put(-505,36){\rotatebox{90}{\large $r=3$}}

  \includegraphics[width=0.19\textwidth,trim={5cm 3.2cm 5cm 3cm}, clip]{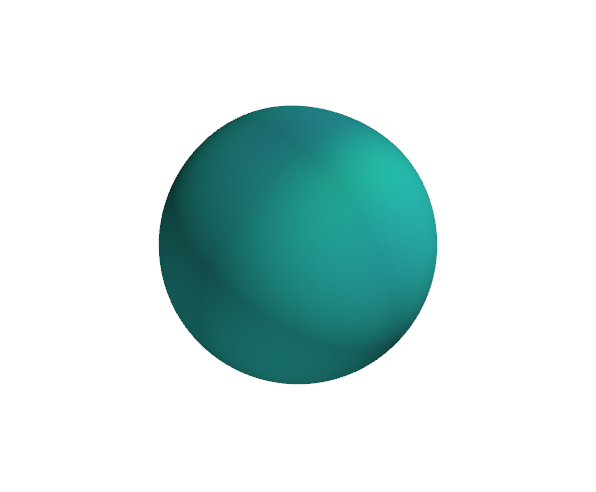}
  \includegraphics[width=0.19\textwidth,trim={5cm 3.2cm 5cm 3cm}, clip]{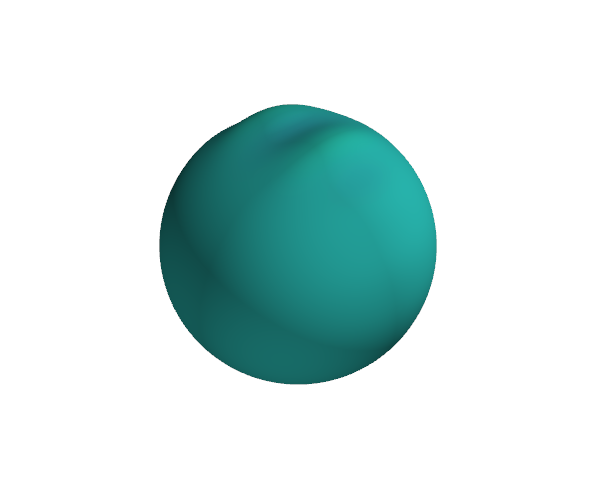}
  \includegraphics[width=0.19\textwidth,trim={5cm 3.2cm 5cm 3cm}, clip]{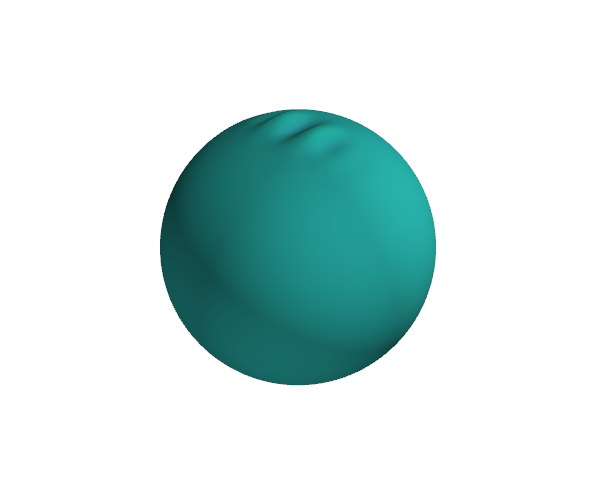}
  \includegraphics[width=0.19\textwidth,trim={5cm 3.2cm 5cm 3cm}, clip]{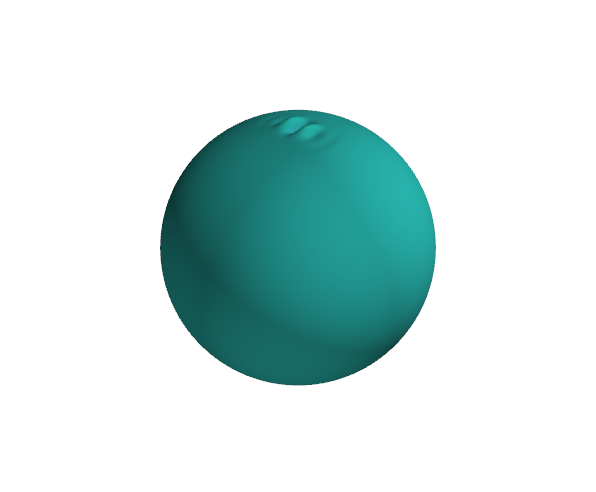}
  \includegraphics[width=0.19\textwidth,trim={5cm 3.2cm 5cm 3cm}, clip]{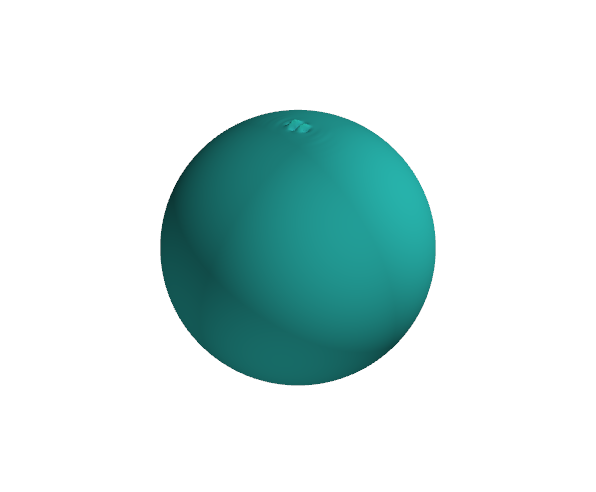}
  \put(-505,36){\rotatebox{90}{\large $r=4$}}

  \caption{Directional ($N=5$) scale-discretised wavelet filters on the ball ($P=L=256$), for angular wavelet scales $j\in[3,7]$ from left to right, and radial wavelet scale $j^\prime = 3$ for radial nodes $r\in[1,4]$. As the angular wavelet scale increases the filter becomes more highly localised, both in pixel and frequency space \citep{mcewen:s2let_localisation}. Equally, as the radial scale increases the localisation along the radial half-line increases. For a given radial scale (as shown here) the energy of a given filter $\Psi(r)$ decays exponentially with $r$ as expected. Notice the clearly elongated structure which gives rise to the desired directional sensitivity.} \label{fig:directional_wavelet_filters_ball}
\end{figure*}

\begin{figure*}
  \centering
  \scalebox{-1}[1]{
    \includegraphics[width=0.19\textwidth,trim={3.5cm 2cm 3.5cm 3cm}, clip]{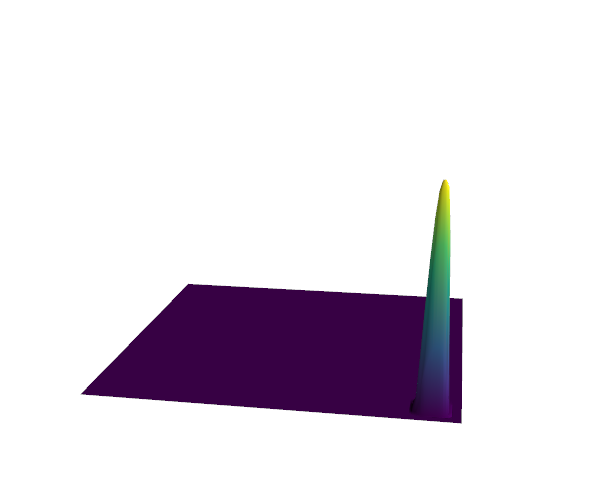}}
  \put(-80,47){\large $\Psi^{3,3}_{\ell 0}$}
  \scalebox{-1}[1]{
    \includegraphics[width=0.19\textwidth,trim={3.5cm 2cm 3.5cm 3cm}, clip]{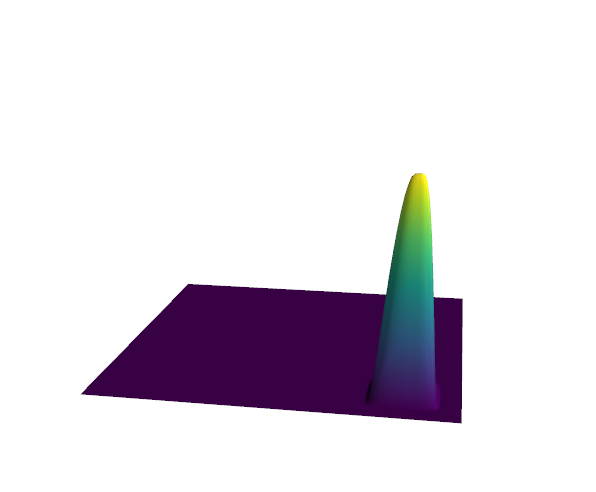}}
  \put(-70,47){\large $\Psi^{4,4}_{\ell 0}$}
  \scalebox{-1}[1]{
    \includegraphics[width=0.19\textwidth,trim={3.5cm 2cm 3.5cm 3cm}, clip]{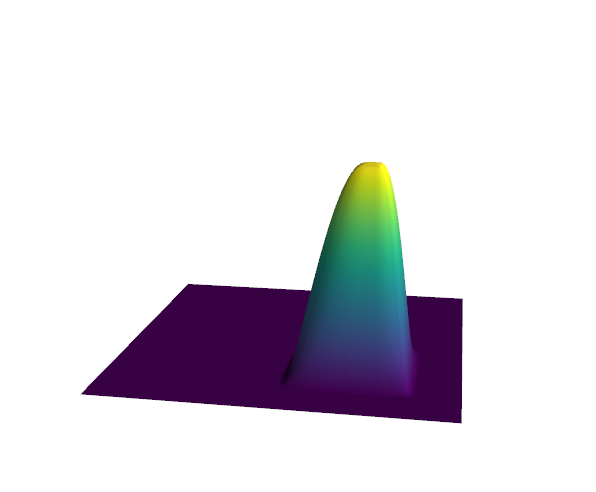}}
  \put(-53,47){\large $\Psi^{5,5}_{\ell 0}$}
  \scalebox{-1}[1]{
    \includegraphics[width=0.19\textwidth,trim={3.5cm 2cm 3.5cm 3cm}, clip]{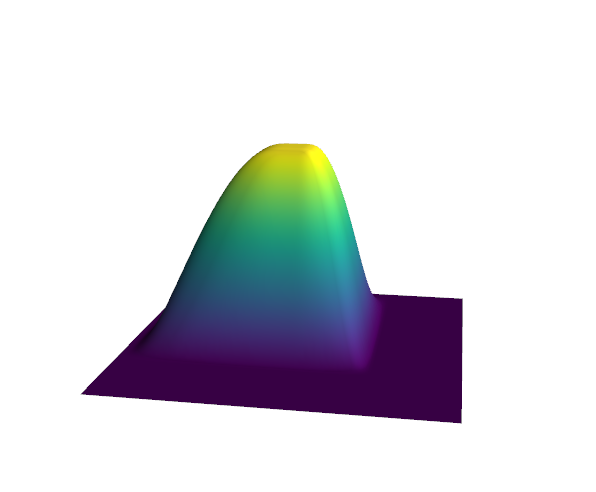}}
  \put(-25,47){\large $\Psi^{6,6}_{\ell 0}$}
  \scalebox{-1}[1]{
    \includegraphics[width=0.19\textwidth,trim={3.5cm 2cm 3.5cm 3cm}, clip]{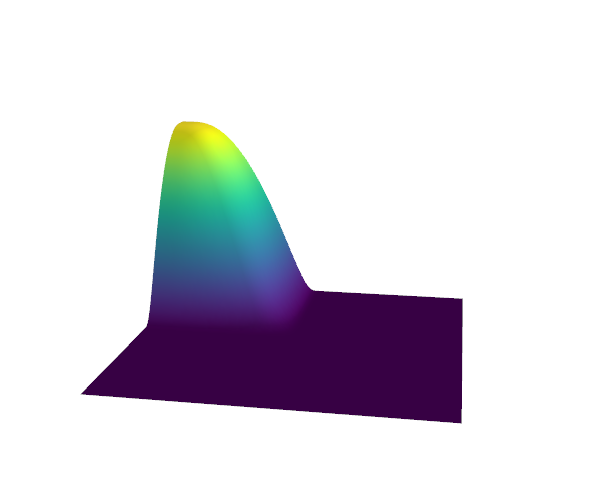}}
  \put(-13,47){\large $\Psi^{7,7}_{\ell 0}$}

  \includegraphics[width=0.19\textwidth,trim={3.5cm 3cm 3.5cm 3cm}, clip]{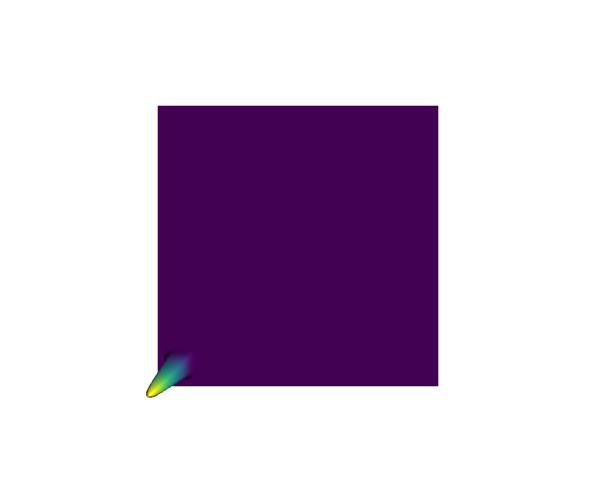}
  \put(-95,37){\large $\ell$}
  \includegraphics[width=0.19\textwidth,trim={3.5cm 3cm 3.5cm 3cm}, clip]{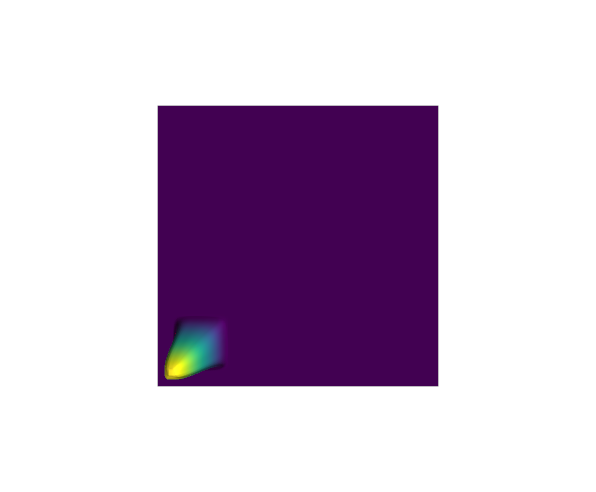}
  \includegraphics[width=0.19\textwidth,trim={3.5cm 3cm 3.5cm 3cm}, clip]{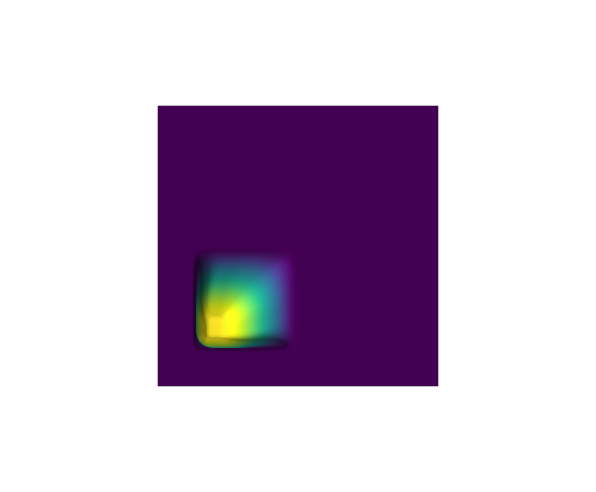}
  \put(-52,-5){\large $p$}
  \includegraphics[width=0.19\textwidth,trim={3.5cm 3cm 3.5cm 3cm}, clip]{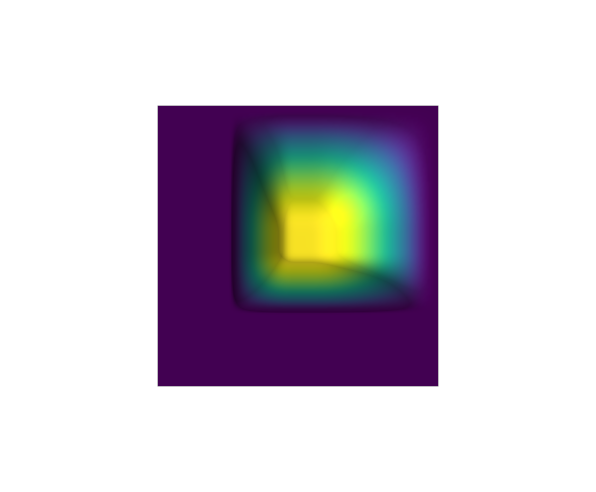}
  \includegraphics[width=0.19\textwidth,trim={3.5cm 3cm 3.5cm 3cm}, clip]{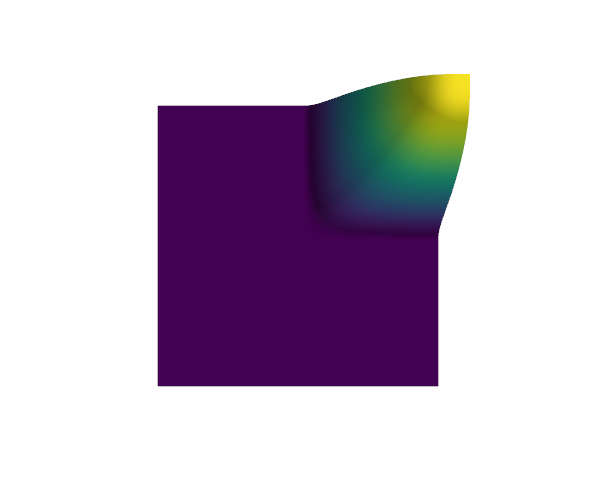}

  \caption{Surface visualisation of a subset of our tiling of Fourier-Laguerre space with infinitely differentiable Cauchy-Schwartz functions (see Appendix \ref{sec:tiling_ball}). Here we present the axisymmetric $(m=0)$ components of our wavelet filters for $j=j^\prime$. Notice that if one were to consider all $j<j^\prime$ and \emph{vice versa} this filtering scheme would span the entire domain.} \label{fig:harmonic_space_tiling}
\end{figure*}
\vspace{10pt}
\subsection{Directional wavelet transform on the ball} \label{sec:b3_wav_cont}
Ball wavelet filters are square integrable functions ${}_s \Psi^{jj^\prime} \in\text{L}^2[\mathbb{B}^3]$ with angular and radial bandlimits $L$ and $P$ respectively. They are designed to exhibit strong localisation properties in both spherical-Laguerre and spatial domain \citep{mcewen:s2let_localisation}, as illustrated in Figure \ref{fig:directional_wavelet_filters_ball}. Here we further introduce $j^\prime \in \mathbb{N} \leq J^\prime$ to denote radial wavelet scale (see Appendix \ref{sec:wav_construct} for further details). As in the spherical case, these wavelet functions form an overcomplete dictionary into which spin-$s$ functions ${}_sf \in \text{L}^2[\mathbb{B}^3]$ may be decomposed through the overloaded directional convolution
\begin{equation}
  W^{{}_s\Psi^{jj^\prime}}(h)
  = ({}_sf \circledast {}_s\Psi^{jj^\prime})(h)
  = \langle {}_sf, \mathcal{L}_{h} \, {}_s\Psi^{jj^\prime} \rangle,
\end{equation}
which from Equation \ref{eq:wig_lag_conv_rep} is represented in Wigner-Laguerre space by the expression
\begin{equation} \label{eq:wig_lag_wav_coeffs}
  (W^{{}_s\Psi^{jj^\prime}})^{\ell}_{mnp}
  = \frac{8\pi^2}{2\ell+1} \:
  {}_s\hat{f}_{\ell m p} \:
  {}_s\Psi^{jj^\prime*}_{\ell n p}.
\end{equation}
To capture low frequency information an axisymmetric scaling function ${}_s\Phi \in \text{L}^2[\mathbb{B}^3]$ is introduced
\begin{equation}
  W^{{}_s\Phi}(b)
  = ({}_sf \odot {}_s\Phi)(b)
  = \langle {}_sf, \mathcal{L}_{b} \, {}_s\Phi \rangle,
\end{equation}
where $\mathcal{L}_{b}$ is the axisymmetric simplification of $\mathcal{L}_{h}$. The spherical-Laguerre representation of the scaling coefficients is given as
\begin{equation} \label{eq:wig_lag_scal_coeffs}
  (W^{{}_s\Phi})_{\ell m p} = \sqrt{\frac{4\pi}{2\ell +1}} {}_sf_{\ell m p} \: {}_s\Phi^*_{\ell 0 p}.
\end{equation}
On the ball, the wavelet admissibility condition as presented in \citet{leistedt:flaglets_spin} reads
\begin{equation}
  \frac{4\pi}{2\ell+1}| {}_s\Phi_{\ell 0 p}|^2
  + \frac{8\pi^2}{2\ell+1}
  \sum_{mjj^\prime}
  |{}_s\Psi^{jj^\prime}_{\ell m p}|^2 = 1 \quad \forall \, \ell, p,
\end{equation}
which, if satisfied, permits exact synthesis by
\begin{align}
  {}_sf
   & = \int_{\mathbb{B}^3} \text{d}\Omega(b) W^{{}_s\Phi}(b)(\mathcal{L}_{b}\Phi)(b) \nonumber                                                                                          \\
   & + \sum_{j,j^\prime=0}^{JJ^\prime} \int_{\mathbb{H}^4} \text{d}\Omega(h) W^{{}_s\Psi^{jj^\prime}}(h)(\mathcal{L}_h \, W^{{}_s\Psi^{jj^\prime}})(b). \label{eq:inverse_ball_wavelet}
\end{align}
This transform can be extremely expensive to evaluate, however noticing that each wavelet scale $jj^\prime$ has compact support can dramatically reduce the total number of calculations required. Such an acceleration is exploited in what is referred to as a multiresolution algorithm.

\subsection{Multiresolution Algorithms} \label{sec:multiscale_algos}
For the development of multiresolution algorithms it is critical to first note that a given wavelet scale has strictly compact support over finitely many harmonic degrees $\ell$, and polynomials $p$, as shown in Figure \ref{fig:harmonic_space_tiling}. Therefore, both the expressions for the forward and inverse wavelet transforms, on both the sphere and ball, may be performed at varying resolutions, without loss of information.

Specifically, the wavelet ${}_s\Psi^j$ at scale $j$ is only non-zero over the finite interval of harmonic degrees
\begin{equation}
  \ell \in \Big [ \: \lfloor\:\lambda^{j-1}\:\rfloor,\: \lceil\: \lambda^{j+1}\: \rceil \: \Big ],
\end{equation}
where $\lfloor\cdot\rfloor$ and $\lceil\cdot\rceil$ are the floor and ceiling functions respectively, and $\lambda \in \mathbb{N} > 1$ is the dilation parameter of the wavelets outlined in Appendix \ref{sec:wav_construct}. As the computational time of a given harmonic degree $\ell$ scales as $\mathcal{O}(\ell^3)$, the overall complexity of the directional wavelet transform is dominated by the highest degrees. It then follows that only the highest two wavelet scales have non-zero support near the harmonic bandlimit $\ell \sim L$, hence the overall complexity is effectively that of this single scale.

Precisely the same argument may be applied in the radial direction, wherein a given radial scale $j^\prime$ has compact support over the finite interval
\begin{equation}
  p \in \Big [ \: \lfloor\:\nu^{j^\prime-1}\:\rfloor,\: \lceil\: \nu^{j^\prime+1}\: \rceil \: \Big ],
\end{equation}
Consequently, multiresolution algorithms designed to evaluate the aforementioned wavelet transforms can be accelerated by a factor of $J$ and $J \times J^\prime$ over the sphere and ball respectively, without loss of information.

\section{The \texttt{S2WAV} library} \label{sec:s2wav}
Wavelet transforms have demonstrated utility in various emergent technologies, whether this be embedding wavelet filters directly within machine learning models \citep{huang:2017,liu:2019} or through multiscale conditioning \citep{guth:2022}. In any case, to encorporate such techniques within modern machine learning technologies requires that the wavelet transforms in question are differentiable, so as to facilitate the back-propagation of gradient information. Furthermore, a primary advantage of such technologies is their high throughput, potentially with real-time evaluation. Therefore the ability to deploy such technologies on hardware accelerators (\emph{e.g.} GPUs/TPUs) is almost mandatory. Wavelet transforms have been developed for Euclidean applications which satisfy the aforementioned remit, however no such transforms exist on the sphere, limiting progress in this area.

In this section we develop and release \texttt{S2WAV} {\href{https://github.com/astro-informatics/s2wav}{\faGithub}}, a professionally developed open-source \texttt{JAX} library \citep{jax:2018:github}, which provides support for the directional wavelet transform on the sphere. Specifically, by leveraging the recently released \texttt{S2FFT} {\href{https://github.com/astro-informatics/s2fft}{\faGithub}} software package \citep{price:s2fft} we provide GPU accelerated and automatically differentiable implementations of the directional scale-discretised wavelet transform outlined in Section \ref{sec:s2_wav_cont}. Building upon novel Wigner $d$-function recursions, \texttt{S2FFT} is designed to be extremely parallelisable, and asymptotically recovers linear scaling across multiple accelerators. Interestingly this results in \texttt{S2FFT} demonstrating an effective linear compute scaling with bandlimit $L$, which is unprecidented. We inherit both this computational scaling and functionality, with \texttt{S2WAV} transforms being efficiently distributable across multiple hardware devices. In what follows we drop spin subscripts for notational brevity.

\subsection{Mathematical Overview}
Explicitly, we are primarily concerned with the efficient evaluation of two transforms corresponding to the forward (analysis) and inverse (synthesis) wavelet transform over $\mathbb{S}^2$. Consider the Wigner space representation of the wavelet coefficients given in Equation \ref{eq:wigner_wav_coeffs}. Introducing $\mathbf{D}^{-1}$ as an operator which applies the inverse Wigner transform in Equation \ref{eq:inverse_wigner_transform} to each scale $j$ and $\mathbf{Y}$ as the forward spin spherical harmonic transform in Equation \ref{eq:fswsh} the wavelet transform of a function $f\in\text{L}^2[\mathbb{S}^2]$ may be written as
    \begin{equation} \label{eq:analysis_s2}
      W^{\Psi^j}
      = \mathbf{D}^{-1} \:
      \mathbf{N} \:
      \mathbf{\Psi}^j \:
      \mathbf{Y}f,
    \end{equation}
    where $\mathbf{\Psi}^j$, with normalised entries given by $\Psi^{j*}_{\ell n}$, is an operator which applies the tensor outer product of $\hat{f}$ with each scale $j$ of a given wavelet dictionary $\{\Psi^j\}$, \emph{i.e.} the directional convolution on the sphere defined in Equation \ref{eq:wigner_wav_coeffs}, and $\mathbf{N}$ applies the normalisation $ 8\pi^2 / (2\ell+1)$. From Equation \ref{eq:harmonic_scal_coeffs} the scaling coefficients are straightforwardly given by
\begin{equation}
  W^{\Phi} =
  \mathbf{Y}^{-1} \:
  \mathbf{\Phi} \:
  \mathbf{Y}f,
\end{equation}
where $\mathbf{\Phi}$, with entries given by $\sqrt{4\pi/(2\ell +1)} \Phi^*_{\ell 0}$, is an operator which denotes the inner product of $\hat{f}$ with a scaling dictionary $\Phi$, \emph{i.e.} the axisymmetric convolution.

Adopting analogous operators, the wavelet synthesis transform presented in Equation \ref{eq:s2_cont_synthesis} may be written as
    \begin{equation} \label{eq:synthesis_s2}
      f = \mathbf{Y}^{-1} \big (
      \mathbf{\Phi}\mathbf{Y} W^{\Phi} \:
      +
      \sum_{j}
      \mathbf{\Psi}^{j\dagger} \:
      \mathbf{D} \:
      W^{\Psi^j}
      \big ),
    \end{equation}
    where the summation is a tensor contraction over each scale $j$. Notice that $\mathbf{\Phi}$ remains the same for both transforms as $\Phi^*_{\ell 0} = \Phi_{\ell 0}$ and the normalisation prefactor is the usual harmonic normalisation in both directions (contrast this with the different normalisations for the Wigner transform, which arise due to the normalisation factor appearing in Equation~\ref{eq:inverse_wigner_transform}).

As the spherical harmonic and Wigner transforms are provided by \texttt{S2FFT}, we need only efficiently compute the necessary tensor operations whilst integrating the multiscale acceleration method outlined in Section \ref{sec:multiscale_algos}.

\subsection{Precomputed Components} \label{sec:s2_precomp}
From Equations \ref{eq:analysis_s2} and \ref{eq:synthesis_s2} it is apparent that the wavelet and scaling filters, $\Psi$ and $\Phi$ respectively, may be calculated and cached for future use rather than evaluated on the fly. In a multiresolution framework, the memory complexity associated with each of these arrays is $\mathcal{O}(NL)$ and $\mathcal{O}(L)$ respectively, which is extremely small. Though perhaps somewhat clunky, this avoids unnecessary compute and potential memory issues due to sequential reallocation of memory during \emph{e.g.} optimisation or training.

Additionally, we configure our transforms to support the precompute functionality within \texttt{S2FFT}. Specifically, this optional acceleration precomputes the real polar-d functions necessary to evaluate forward spin spherical harmonic, and by extension Wigner, transforms. Though this can provide extremely fast transforms, it comes with an $\mathcal{O}(NL^3)$ memory overhead, which limits the resolution at which these transforms may be applied \citep[\emph{e.g.}][]{cobb:efficient_generalized_s2cnn}.

\subsection{\texttt{JAX} Tensor Operations}
Given precomputed wavelet and scaling filters, implementing highly efficient algorithms to evaluate the tensor operations $\mathbf{\Psi}$ and $\mathbf{\Phi}$ is straightforward. By design, \texttt{JAX} in fact provides support for all such operations through \texttt{einsum}, which converts understandable symbolic notation into linear algebraic array operations. This transform is outlined in Algorithm \ref{alg:s2_wavelets_algo} which includes a sketch of the associated code.

As both the spherial harmonic and Wigner transforms provided by \texttt{S2FFT} and the \texttt{einsum} primitives are natively differentiable, the directional wavelet transforms provided by \texttt{S2WAV} also provide automatic differentiation. Our transforms can therefore be straightforwardly integrated within existing frameworks to extend, \emph{e.g.} multiscale conditioned generative models \citep{huang:2017,liu:2019,guth:2022} or scattering covariances \citep{allys:2020:new,mallat:2020:phase} to the spherical setting (Mousset \emph{et al} in prep).

Moreover, we design \texttt{S2WAV} to utilise the single program multiple data (SPMD) functionality of \texttt{S2FFT} to distribute compute across hardware accelerators. Given that the complexity of Equations \ref{eq:analysis_s2} and \ref{eq:synthesis_s2} is dominated by the Wigner transforms, and noting the discussion of \citet{price:s2fft}, with sufficient compute our wavelet transforms asymptotically recover an effective linear time complexity. In the case where a small number of GPU devices are available one should expect to asymptotically recover a further acceleration by the number of devices.

\begin{figure}[t]
  \begin{algorithm}[H]
    \caption{Directional wavelet transform on $\mathbb{S}^2$} \label{alg:s2_wavelets_algo}
    \begin{algorithmic}[0]
      \State \texttt{import s2fft}
      \Comment{{\scriptsize \texttt{JAX} Spherical harmonic transforms}}
      \vspace{2pt}

      \State \texttt{from s2fft import wigner}
      \Comment{{\scriptsize \texttt{JAX} Wigner transforms}}
      \vspace{2pt}

      \State \texttt{from jax.numpy import einsum}
      \Comment{{\scriptsize \texttt{JAX} Tensor operations}}
      \vspace{2pt}

      \Procedure{Analysis wavelet transform}{$f\in\mathbb{S}^2$}:
      \vspace{2pt}

      \State $\hat{f} \leftarrow \mathbf{Y}f$
      \Comment{{\scriptsize \texttt{s2fft.forward($f$,$L$)}}}
      \vspace{2pt}

      \State $\hat{W}^{\Phi} \leftarrow \mathbf{\Phi}\hat{f}$
      \Comment{{\scriptsize \texttt{einsum("lm,l->lm",$\hat{f}$,$\Phi$)}}}
      \vspace{2pt}

      \State $W^{\Phi} \leftarrow \mathbf{Y}^{-1}\hat{W}^{\Phi}$
      \Comment{{\scriptsize \texttt{s2fft.inverse($\hat{W}^{\Phi}$,$L$)}}}
      \vspace{2pt}

      \For{$j \in [0,J]$}
      \vspace{2pt}

      \State $\hat{W}^{\Psi_j} \leftarrow \mathbf{N} \mathbf{\Psi}^j \hat{f}$
      \Comment{{\scriptsize \texttt{einsum("lm,ln->nlm",$\hat{f}$,$\mathbf{N} \Psi^{j*}$)}}}
      \vspace{2pt}

      \State $W^{\Psi_j} \leftarrow \mathbf{D}^{-1}\hat{W}^{\Psi_j}$
      \Comment{{\scriptsize \texttt{wigner.inverse($\hat{W}^{\Psi_j}$,$L$,$N$)}}}
      \vspace{2pt}

      \EndFor
      \State \textbf{return} $\big \lbrace W^{\Psi}, W^{\Phi} \big \rbrace$
      \EndProcedure
      \vspace{2pt}

      \Procedure{Synthesis wavelet transform}{$\big \lbrace W^{\Psi}, W^{\Phi} \big \rbrace$}:
      \vspace{2pt}

      \State $\hat{W}^{\Phi} \leftarrow \mathbf{Y}W^{\Phi}$
      \Comment{{\scriptsize \texttt{s2fft.forward($W^{\Phi}$,$L$)}}}
      \vspace{2pt}

      \State $\hat{f} \leftarrow \mathbf{\Phi}\hat{W^{\Phi}}$
      \Comment{{\scriptsize \texttt{einsum("lm,l->lm",$\hat{W^{\Phi}}$,$\Phi$)}}}
      \vspace{2pt}

      \For{$j \in [0,J]$}
      \vspace{2pt}

      \State $\hat{W}^{\Psi_j} \leftarrow \mathbf{D}W^{\Psi_j}$
      \Comment{{\scriptsize \texttt{wigner.forward($W^{\Psi_j}$,$L$,$N$)}}}
      \vspace{2pt}

      \State $\hat{f} \xleftarrow[]{+}\mathbf{\Psi}^{\dagger} \hat{W}^{\Psi_j}$
      \Comment{{\scriptsize \texttt{einsum("nlm,ln->lm",$\hat{W}^{\Psi_j}$,$\Psi^{j}$)}}}
      \vspace{2pt}

      \EndFor

      \State $f \leftarrow \mathbf{Y}^{-1}\hat{f}$
      \Comment{{\scriptsize \texttt{s2fft.inverse($\hat{f}$,$L$)}}}
      \vspace{2pt}
      \State \textbf{return} $f$
      \EndProcedure
    \end{algorithmic}
  \end{algorithm}
  \vspace{-5pt}
\end{figure}

\subsection{Numerical Validation} \label{sec:protocol}
We benchmark our directional wavelet transforms against the existing \texttt{C} alternative \texttt{S2LET} {\href{https://github.com/astro-informatics/s2let}{\faGithub}} \citep{leistedt:s2let_axisym,mcewen:s2let_spin}. Our protocol is straightforward: generate a random bandlimited signal $\hat{f}$; map this to a random bandlimited function on the sphere $f\leftarrow\mathbf{Y}^{-1}\hat{f}$; apply the forward wavelet transform followed by the inverse transform to recover $f^\prime$; map this function onto its harmonic coefficients $\hat{f}^\prime \leftarrow \mathbf{Y}f^\prime$; and evaluate both the round-trip time and relative error $\mathbb{E}(|\hat{f} - \hat{f}^\prime|)$.

\texttt{S2LET} transforms were executed on a multithreaded Xeon(R) E5-2650L v3 dedicated CPU and \texttt{S2WAV} transforms were executed on both a single and three NVIDIA A100 GPUs. Further testing over many more GPUs has been left for future development due to computational constraints. In addition, we provide benchmark results for both operating modalities discussed in Section \ref{sec:s2_precomp}; wherein real polar $d$-functions are evaluated on-the-fly or simply precomputed.

The results of this benchmarking are presented for recursive and precompute transforms in Table \ref{tab:sphere_timings_recursive}. As expected, when distributing across three GPUs we asymptotically recover an additional factor of three acceleration, reaching as high as $100\times$ and $300\times$ faster than existing \texttt{S2LET} transforms, for recursive and precompute respectively. In all cases our transforms are exact to 64-bit machine precision. Note that here we adopt McEwen-Wiaux sampling which affords a sampling theorem \citep{mcewen:fssht,mcewen:so3} and is theoretically exact. We also provide support for HEALPix sampling \citep{gorski:2005} which does not support a sampling theorem, and therefore produces approximate transforms.

\begin{table*}[htbp]
  \newcolumntype{C}[1]{>{\centering\let\newline\\\arraybackslash\hspace{0pt}}m{#1}}
  \centering 
  \begin{tabular}{C{0.05\textwidth} C{0.1\textwidth} C{0.11\textwidth} C{0.08\textwidth} C{0.08\textwidth} C{0.1\textwidth} C{0.11\textwidth} C{0.08\textwidth}}
    \toprule
    \midrule
           &
           & \multicolumn{4}{c}{On-the-fly transform}
           & \multicolumn{2}{c}{Precompute transform} \\
    \cmidrule(lr){7-8}
    \cmidrule(lr){3-6}
    $L$
           & Time \texttt{S2LET}
           & Time \texttt{S2WAV}
           & Speed-Up 1 $\times$ GPU
           & Speed-Up 3 $\times$ GPU
           & 3 $\times$ GPU Ratio
           & Time \texttt{S2WAV}
           & Speed-Up 1 $\times$ GPU                  \\ \midrule

    $8$    & \hspace{5pt}$1.8 \: \times10^{-1}$
           & --
           & --
           & --
           & --
           & \hspace{5pt}$1.4 \: \times10^{-1}$
           & 1.3                                      \\[0.1cm]

    $16$   & $1.3 \: \times10^{0}$
           & --
           & --
           & --
           & --
           & \hspace{5pt}$1.6 \: \times10^{-1}$
           & 8.1                                      \\[0.1cm]

    $32$   & $9.5 \: \times10^{0}$
           & --
           & --
           & --
           & --
           & \hspace{5pt}$2.2 \: \times10^{-1}$
           & 43.2                                     \\[0.1cm]

    $64$   & $3.3 \: \times10^{1}$
           & $7.9 \: \times10^{1}$
           & $0.4$
           & $0.7$
           & $1.8$
           & \hspace{5pt}$4.6 \: \times10^{-1}$
           & 71.7                                     \\[0.1cm]

    $128$  & $2.1 \: \times10^{2}$
           & $1.5 \: \times10^{2}$
           & $1.4$
           & $2.8$
           & $2.0$
           & $1.5 \: \times10^{0}$
           & 140                                      \\[0.1cm]

    $256$  & $1.8 \: \times10^{3}$
           & $3.0 \: \times10^{2}$
           & $6.0$
           & $13$
           & $2.2$
           & $6.0 \: \times10^{0}$
           & 300                                      \\[0.1cm]

    $512$  & $1.3 \: \times10^{4}$
           & $9.4 \: \times10^{2}$
           & $14$
           & $32$
           & $2.3$
           & --
           & --                                       \\[0.1cm]

    $1024$ & $9.8 \: \times10^{4}$
           & $4.3 \: \times10^{3}$
           & $23$
           & $58$
           & $2.5$
           & --
           & --                                       \\[0.1cm]

    $2048$ & $1.1 \: \times10^{6}$
           & $3.0 \: \times10^{4}$
           & $37$
           & $92$
           & $2.5$
           & --
           & --                                       \\[0.1cm]
    \midrule
    \bottomrule
  \end{tabular}
  \caption{\footnotesize Numerical validation of the recursive (on-the-fly) and precompute \texttt{S2WAV} directional wavelet transform on the sphere against the existing \texttt{S2LET} package. Note that all timings are quoted in milliseconds. The precompute mode of \texttt{S2WAV} caches elements of the real polar $d$-functions necessary to evaluate the spin spherical harmonics transforms. Therefore, though this greatly increases throughput, the peak memory overhead scales as $\mathcal{O}(L^3)$ which limits the maximum operational bandlimit. Here we consider a single fixed azimuthal bandlimit $N=5$ which corresponds to $9$ rotations of our wavelet filters within each local tangent plane; for the majority of applications this is more than sufficient to recover almost all directional structure. \texttt{S2LET} functions were executed on a multithreaded Xeon(R) E5-2650L v3 CPU and \texttt{S2WAV} transforms were evaluated on a single and collection of three NVIDIA A100 GPUs. For low bandlimits $L$ communication and GPU memory allocation costs slow down our transform, however in such cases compute is rarely an issue. Furthermore at low to moderate resolutions a precompute approach may be adopted, which can provide up to $\sim 300 \times$ faster. In higher resolution cases our transforms become up to $\sim 100 \times$ faster, whilst retaining 64-bit machine precision.} \label{tab:sphere_timings_recursive}
  \vspace{4pt}
  \hrule
\end{table*}

\section{The \texttt{S2BALL} library} \label{sec:baller}
As discussed in Section \ref{sec:s2wav}, it has been demonstrated that various modern machine learning technologies may be enhanced by the incorporation of wavelet representations; with the added caveat that the associated wavelet transforms must be differentiable and, ideally, deployable on hardware accelerators. Hardware acceleration is even more critical when considering functions on the ball. In short, this is because wavelet transforms over this space are prohibitively expensive. In addition to the aforementioned contributions, extending spherical wavelet dictionaries radially opens up many salient scientific applications, particularly in the study of geophysics \citep{simons:2011,marignier:2020:probability} and molecular modelling \citep{boomsma:2017:spherical,jumper:alphafold}.

To this end, we develop and release \texttt{S2BALL} {\href{https://github.com/astro-informatics/baller}{\faGithub}}, a highly optimised open-source \texttt{JAX} library which provides support for the directional wavelet transform on the ball. Specifically, we provide GPU accelerated and automatically differentiable implementations of the Wigner-Laguerre wavelet transform outlined in Section \ref{sec:b3_wav_cont}. In this case, we do not integrate the \texttt{S2FFT} {\href{https://github.com/astro-informatics/s2fft}{\faGithub}} package for spin spherical harmonic and Wigner transforms, instead providing bespoke implementations for computational efficiency, as will become apparent.

\subsection{Mathematical Overview}
Explicitly, this package is designed to efficiently evaluate the forward (analysis) and inverse (synthesis) directional wavelet transform of $L,P$ bandlimited functions $f \in \text{L}^2[\mathbb{B}^3]$. First, consider the Wigner-Laguerre space representation of the wavelet coefficeints given in Equation \ref{eq:wig_lag_wav_coeffs}. Recalling our definition for the inverse Wigner-Laguerre given in Equation \ref{eq:wiglag_inverse} it is clear that the pixel-space representation of scale $\lbrace j,j^\prime \rbrace$ is given by
\begin{equation} \label{eq:wavelet_part1}
  W^{\Psi^{jj^\prime}}(h)
  = \sum_{p, \ell\,\in\,\mathbb{N}} \: \frac{2\ell+1}{8\pi^2} \,
  \sum_{|m,n|\,\leq\,\ell} \,
  (W^{\Psi^{jj^\prime}})^{\ell}_{mnp} \:
  Q^{\ell*}_{mnp}(h).
\end{equation}
Introducing the operators $\mathbf{Z}$ to denote the forward spherical-Laguerre transforms defined in Equation \ref{eq:forward_spherical_laguerre}, and introducing $\mathbf{Q}^{-1}$ to denote the inverse Wigner-Laguerre transform defined in Equation \ref{eq:wiglag_inverse}, one may recast Equation \ref{eq:wavelet_part1} to read
    \begin{equation} \label{eq:forward_ball_wavelet}
      W^{\Psi^{jj^\prime}}
      = \mathbf{Q}^{-1} \:
      \mathbf{N}
      \mathbf{\Psi}^{jj^\prime} \:
      \mathbf{Z}f,
    \end{equation}
    where $\mathbf{\Psi}^{jj^\prime}$ is an operator which applies a scale-wise tensor outer product of $\hat{f}$ with each scale $\lbrace j,j^\prime \rbrace$ of a given wavelet dictionary $\{\Psi^{j,j^\prime}\}$, \emph{i.e.} the directional convolution on the ball defined in Equation \ref{eq:wig_lag_wav_coeffs}. Adopting similar operators, from Equation \ref{eq:wig_lag_scal_coeffs} the scaling coefficients are given by
\begin{equation}
  W^{\Phi} =
  \mathbf{Z}^{-1} \:
  \mathbf{\Phi} \:
  \mathbf{Z}f,
\end{equation}
where $\mathbf{\Phi}$ is an operator which applies the tensor inner product with scaling dictionary $\Phi$, \emph{i.e.} the axisymmetric convolution. Finally, the continuous synthesis transform presented in Equation \ref{eq:inverse_ball_wavelet} may be written as
    \begin{equation} \label{eq:synthesis_b3}
      f = \mathbf{Z}^{-1} \big (
      \mathbf{\Phi}\mathbf{Z} W^{\Phi} \:
      + \sum_{jj^\prime}
      \mathbf{\Psi}^{jj^\prime\dagger}
      \mathbf{Q} \:
      W^{\Psi^{jj^\prime}}
      \big ),
    \end{equation}
    where the summation performs a tensor contraction over each scale $\lbrace j,j^\prime \rbrace$.
\vspace{15pt}
\subsection{Precomputed Components} \label{sec:b3_precomp}
Na\"ive matrix representations of $\mathbf{Q}$ and $\mathbf{Z}$ can unsurpsingly become rather large. Noting that the memory complexity of a function $f\in\text{L}^2[\mathbb{H}^4]$ scales as $\mathcal{O}(NPL^2)$, and that $P$ typically weakly scales as $L$, we cannot hope for better than $\sim\mathcal{O}(NL^3)$ memory requirements. With some care, the memory complexity for matrix representations of $\mathbf{Q}$ and $\mathbf{Z}$ can be reduced to $\mathcal{O}(NL^3)$.

In the same way that the Wigner-Laguerre basis functions can be decomposed into the orthogonal Laguerre and Wigner basis functions, the Wigner-Laguerre transform can be expanded as $\mathbf{Q} = \mathbf{K}\mathbf{D}$ where $\mathbf{K}$ represents the forward Laguerre transform defined in Equation \ref{eq:flag}. Furthermore, by decomposing the Wigner-$D$ functions in terms of real polar $d$-functions
\begin{equation}
  D^{\ell}_{mn}(\alpha, \beta, \gamma)
  = d^{\ell}_{mn}(\beta)
  e^{-i(n\gamma+m\alpha)}
\end{equation}
the Wigner transform defined in Equation \ref{eq:forward_wigner_transform} reduces to
\begin{equation}
  \hat{f}
  = \int_{\beta} d^{\ell}_{mn}(\beta)
  \underbrace{
    \int_{\alpha, \gamma} \text{d}\Omega(\rho) \: f(\rho) \: e^{-i(n\gamma+m\alpha)}
  }_{\text{Fourier Transform}}.
\end{equation}
In the discrete setting one may evaluate this double integral as a 2-dimensional fast Fourier transform \citep{cooley:1965}, which require minimal memory and are famously extremely efficient with $\mathcal{O}(N^2\log N)$ complexity. The integral over $\beta$ is effectively a projection onto the real polar $d$-functions, which are often computed recursively \citep{varshalovich:1989}. However, to store all necessary $d^{\ell}_{mn}(\beta)$ requires $\mathcal{O}(NL^3)$ memory which is equivalent to the memory required to store a single signal on the ball, as outlined above. Therefore, for our purposes we suffer no additional memory complexity from simply precomputing and caching these matrices, with the enormous upside of theoretically optimal speed at runtime.

For completeness, consider the Laguerre transform $\mathbf{K}$ defined in Equation \ref{eq:flag}. This is simply a projection onto the Laguerre polynomials $K_p(r)$ with memory complexity $\sim\mathcal{O}(P^2) \leq \mathcal{O}(L^2)$ and therefore negligible. Hence, we may also precompute and store all $K_p(r)$ for efficiency. Finally, we must consider the memory required to store the wavelet $\Psi$ and scaling $\Phi$ filters. The scaling filter $\Phi$ comes with memory complexity $\mathcal{O}(PL) \leq \mathcal{O}(L^2)$ which is negligible. In contrast, the wavelet filters $\Psi$ exhibit memory complexity $\mathcal{O}(JJ^\prime NPL)$ which can nominally be rather large. Adopting the multiresolution scheme discussed in Section \ref{sec:multiscale_algos} only the highest scale $\lbrace J, J^\prime \rbrace$ need be stored at full resolution, and thus the complexity becomes $\mathcal{O}(NPL) \leq \mathcal{O}(L^3)$. We are therefore free to compute and cache all necessary tensors offline with at most $\mathcal{O}(NL^3)$ memory, which we subsequently use to execute transforms extremely efficiently at runtime.

Identical arguments hold for the spherical-Laguerre transform and their corresponding inverse transforms. In \texttt{S2BALL} we default to this operational modality, which has the added benefit of automatically providing the adjoint wavelet transformations needed for optimisation based algorithms.

\subsection{\texttt{JAX} Tensor Operations}
As discussed above, we can precompute effectively all intermediate tensors. Consequently, implementing extremely distributed operations is very straightforward. The entire ball wavelet transform can be written using the aforementioned \texttt{einsum} and the \texttt{JAX} fast Fourier transform API. This transform is outlined in Algorithm \ref{alg:b3_wavelets_algo} which includes a sketch of the associated code.

In this case providing cross GPU distribution through SPMD is not necessary. At low resolutions the additional communication overhead outweighs the potential acceleration. At high resolutions this balance may tip in the favour of distribution across multiple GPUs, however long before this point the memory overhead becomes untenable. In theory, one could shard precomputed tensors across multiple (potentially many) GPU devices, however we leave this for future work.

\begin{figure}[t]
  \begin{algorithm}[H]
    \caption{Directional wavelet transform on $\mathbb{B}^3$} \label{alg:b3_wavelets_algo}
    \begin{algorithmic}[0]
      \State \texttt{from baller import laguerre}
      \vspace{2pt}
      \State \texttt{from baller import wigner\_laguerre as wlaguerre}
      \vspace{2pt}
      \State \texttt{from jax.numpy import einsum}
      \vspace{2pt}

      \Procedure{Analysis wavelet transform}{$f\in\mathbb{B}^3$}:
      \vspace{2pt}

      \State $\hat{f} \leftarrow \mathbf{Z}f$
      \Comment{{\scriptsize \texttt{laguerre.forward($f$,$L$,$P$)}}}
      \vspace{2pt}

      \State $\hat{W}^{\Phi} \leftarrow \mathbf{\Phi}\hat{f}$
      \Comment{{\scriptsize \texttt{einsum("plm,pl->plm",$\hat{f}$,$\Phi$)}}}
      \vspace{2pt}

      \State $W^{\Phi} \leftarrow \mathbf{Z}^{-1}\hat{W}^{\Phi}$
      \Comment{{\scriptsize \texttt{laguerre.inverse($\hat{W}^{\Phi}$,$L$,$P$)}}}
      \vspace{2pt}

      \For{$j \in [0,J]$ and $j^\prime \in [0,J^\prime]$}
      \vspace{2pt}
      \State $\hat{W}^{\Psi^{jj^\prime}} \leftarrow \mathbf{N} \mathbf{\Psi}^{jj^\prime} \hat{f}$
      \Comment{{\scriptsize \texttt{einsum("plm,pln->pnlm",$\hat{f}$,$\mathbf{N}\Psi^{jj^\prime*}$)}}}
      \vspace{2pt}

      \State $W^{\Psi^{jj^\prime}} \leftarrow \mathbf{Q}^{-1}\hat{W}^{\Psi^{jj^\prime}}$
      \Comment{{\scriptsize \texttt{wlaguerre.inverse($\hat{W}^{\Psi^{jj^\prime}}\hspace{-5pt},L,P,N$)}}}
      \vspace{2pt}
      \EndFor
      \State \textbf{return} $\big \lbrace W^{\Psi}, W^{\Phi} \big \rbrace$
      \EndProcedure
      \vspace{2pt}

      \Procedure{Synthesis wavelet transform}{$\big \lbrace W^{\Psi}, W^{\Phi} \big \rbrace$}:
      \vspace{2pt}

      \State $\hat{W^{\Phi}} \leftarrow \mathbf{Z}W^{\Phi}$
      \Comment{{\scriptsize \texttt{laguerre.forward($W^{\Phi}$,$L$,$P$)}}}
      \vspace{2pt}

      \State $\hat{f} \leftarrow \mathbf{\Phi}\hat{W^{\Phi}}$
      \Comment{{\scriptsize \texttt{einsum("plm,pl->plm",$\hat{W^{\Phi}}$,$\Phi$)}}}
      \vspace{2pt}

      \For{$j \in [0,J]$ and $j^\prime \in [0,J^\prime]$}
      \vspace{2pt}

      \State $\hat{W}^{\Psi^{jj^\prime}} \leftarrow \mathbf{Q}W^{\Psi^{jj^\prime}}$
      \Comment{{\scriptsize \texttt{wlaguerre.forward($W^{\Psi^{jj^\prime}}$,$L$,$P$,$N$)}}}
      \vspace{2pt}

      \State $\hat{f} \xleftarrow[]{+}\mathbf{\Psi}^{\dagger} \hat{W}^{\Psi^{jj^\prime}}$
      \Comment{{\scriptsize \texttt{einsum("pnlm,pln->plm",$\hat{W}^{\Psi^{jj^\prime}}$,$\Psi^{jj^\prime}$)}}}
      \vspace{2pt}

      \EndFor

      \State $f \leftarrow \mathbf{Z}^{-1}\hat{f}$
      \Comment{{\scriptsize \texttt{laguerre.inverse($\hat{f}$,$L$,$P$)}}}
      \vspace{2pt}
      \State \textbf{return} $f$

      \EndProcedure

    \end{algorithmic}
  \end{algorithm}
  \vspace{-15pt}
\end{figure}

\subsection{Numerical Validation}
We benchmark our directional wavelet transform against the existing \texttt{C} alternative \texttt{FLAGLET} {\href{https://github.com/astro-informatics/src_flaglet}{\faGithub}} \citep{leistedt:flaglets,mcewen:flaglets_sampta}. Our protocol is the same as outlined in Section \ref{sec:protocol}, with spherical harmonic transforms $\mathbf{Y}$ interchanged with spherical-Laguerre transforms $\mathbf{Z}$. \texttt{FLAGLET} transforms were executed on a multithreaded Xeon(R) E5-2650L v3 dedicated CPU and \texttt{S2BALL} transforms were executed on a single NVIDIA A100 GPU.

During these tests we restrict ourselves to azimuthal bandlimit $N=1$ for simplicity, and maintain the most general setting of $P=L$ radially. It should be noted that in practice one may work with different radial and angular resolutions, which are often dictated by the data at hand. The results of this benchmark are presented in Table \ref{tab:ball_timings}, which includes both round-trip precision and timings. In every case we recover 64-bit machine precision. For low bandlimits $L\sim8$ our ball wavelet transform is an order of magnitude faster than their \texttt{C} counterparts. At higher resolutions this acceleration increases dramatically, peaking at $21800$ times faster for $L=P=256$. This substantial acceleration opens up the possibility of computationally expensive statistical methods such as Bayesian sampling methods, with direct applications in \emph{e.g.} geophysical imaging \citep{marignier:s2proxmcmc}.

An additional factor worth highlighting is somewhat hidden within the wavelet transform. The spherical- and Wigner-Laguerre transforms we implement and use to construct the \texttt{S2BALL} wavelet transforms are, in and of themselves, interesting and of potential use. In much the same way that spherical harmonic and Wigner transforms can be seen as generalised Fourier transforms (GFFTs) over $\mathbb{S}^2$ and SO(3) respectively, the spherical- and Wigner-Laguerre transforms can be seen to be GFFTs on $\mathbb{B}^3$ and $\mathbb{H}^4$ respectively. In addition to wavelets, \texttt{S2BALL} provides \texttt{JAX} functions with which one may readily evaluate these transforms.

Efficient, differentiable, and exact Fourier based convolution algorithms on $\mathbb{B}^3$ and $\mathbb{H}^4$ present an exciting opportunity to develop rotational and radially translational equivariant networks on 3-dimensional spaces; a natural progression of equivariant learning on the sphere \citep[see \emph{e.g.}][]{cohen:2018:spherical,esteves:2020:spin, bronstein:2021:geometric, cobb:efficient_generalized_s2cnn,ocampo:2023:scalable}. Not only do such networks provide state-of-the-art performance but they are much more data-efficient, which is a key consideration for modern astrophysics.

\begin{table}[htbp]
  \newcolumntype{C}[1]{>{\centering\let\newline\\\arraybackslash\hspace{0pt}}m{#1}}
  \centering 
  \begin{tabular}{c C{0.15\textwidth} C{0.15\textwidth} C{0.065\textwidth}}
    \toprule
    \midrule
    \multicolumn{4}{c}{Validation of directional ball wavelet $(P=L)$}        \\
    \midrule

    $L$   & Time \texttt{FLAGLET}              & Time \texttt{S2BALL} & Ratio \\

    \midrule
    $8$   & $7.0 \: \times10^{0}$
          & \hspace{5pt}$7.5 \: \times10^{-1}$
          & $13$                                                              \\[0.1cm]

    $16$  & $1.0 \: \times10^{2}$
          & \hspace{5pt}$9.5 \: \times10^{-1}$
          & $116$                                                             \\[0.1cm]

    $32$  & $8.3 \: \times10^{2}$
          & $1.3 \: \times10^{0}$
          & $698$                                                             \\[0.1cm]

    $64$  & $1.0 \: \times10^{4}$
          & $3.0 \: \times10^{0}$
          & $3690$                                                            \\[0.1cm]

    $128$ & $1.0 \: \times10^{5}$
          & $1.7 \: \times10^{1}$
          & $6400$                                                            \\[0.1cm]

    $256$ & $3.2 \: \times10^{6}$
          & $1.6 \: \times10^{2}$
          & $21800$                                                           \\[0.1cm]
    \midrule
    \bottomrule
  \end{tabular}
  \caption{\footnotesize Numerical validation of our Wigner-Laguerre wavelet transform against the previous implementation \texttt{FLAGLET}. In each case we consider $N=1$ and maintain a highly general setting where $L=P$. All CPU operations were evaluated on a multithreaded Xeon(R) E5-2650L v3 CPU whereas our transforms were evaluated on a single NVIDIA A100 GPU. The overall complexity scaling of the Wigner-Laguerre transform is $\sim \mathcal{O}(NL^4)$, with an associated memory overhead of $\mathcal{O}(NL^3)$ \emph{i.e.} the memory required to store the highest scale wavelet coefficients a single function $f\in\mathbb{H}^4$. In all cases we are at least an order of magnitude faster, rising as high as $\sim 22,000$ times faster, whilst being exact to machine precision.} \label{tab:ball_timings}
  \vspace{4pt}
  \hrule
\end{table}
\section{Conclusions} \label{sec:conclusions}
In this article we develop novel efficient algorithms for the directional scale-discretised wavelet transforms on the sphere and ball. By construction, our transforms are automatically differentiable and highly distributable both on and across hardware accelerators, \emph{e.g.} GPUs/TPUs. To increase accesibility, and in the spirit of reproducible science, we provide professionally developed open-sourced \texttt{JAX} libraries in both settings.

For the multiscale analysis of functions on the sphere, we publicly release \texttt{S2WAV} {\href{https://github.com/astro-informatics/s2wav}{\faGithub}}. This package provides \texttt{JAX} algorithms for scalable and automatically differentiable directional wavelet transforms on the sphere. These transforms are sampling agnostic, and may be distributed across many GPUs for extreme parallelisation. Through benchmarking we find that our transforms are several orders of magnitude faster than their \texttt{C} counterparts, whilst retaining 64-bit floating point precision.

For the multiscale analysis of functions on the ball, we publicly release \texttt{S2BALL} {\href{https://github.com/astro-informatics/baller}{\faGithub}}. This package provides \texttt{JAX} algorithms for extremely fast but memory intensive directional wavelet transforms on the ball. As part of \texttt{S2BALL}, we provide APIs to evaluate differentaible fast Fourier transforms on the ball and its SO(3) anologue. Our transforms are built on a well-known Laguerre discretisation of the radial half-line, however unlike previous methods our angular components are sampling agnostic. All transforms are automatically differentiable. Through benchmarking we find that our transforms are $\sim 22$ thousand times faster than their \texttt{C} counterparts, whilst retaining 64-bit floating point precision.

The algorithms and software we provide are critical for the fusion of wavelet theory and machine learning on the sphere and ball, unlocking the advantages such a union entails. Wavelet enhanced technologies over Euclidean spaces are already producing state-of-the-art results, \emph{e.g.}\ as embedded representations \citep{huang:2017}, for multiscale conditioning during diffusion-based generative models \citep{guth:2022}, or for equivariant machine learning \citep{mcewen:scattering,ocampo:2023:scalable}. This work provides the tools by which anologous results may be realised on the sphere and ball.

Concomitantly with this work, we are currently developing spherical scattering covariances (an extremely compact wavelet-based representation) which effectively encodes complex non-Gaussian structure and from which realistic cosmological fields may readily be generated (\citet{mousset:scatcov} in prep), which relies directly on this current work. In future work we will encorporate the results of this work to enhance diffusion models on the sphere, with a plethora of applications including weather and climate prediction tasks. As differentiable programming and hardware acceleration grows in popularity throughout the scientific community, it is pertinent that our software tools are modernised. In this work we do just this, infusing the previous generation of multiscale analysis tools with characteristics necessary for integration to truly next generation technologies.

\section*{Acknowledgements}
M.A.P. and J.D.M. are supported by EPSRC (grant number EP/W007673/1). A.P. is supported by the UCL Centre for Doctoral Training in Data Intensive Science (STFC grant number ST/W00674X/1). J.W. is supported by a Science \& Technology Facilities Council PhD Studentship (STFC grant number EP/T517793/1).

\section*{Contribution Statement}
Author contributions are specified below, following the Contributor Roles Taxonomy (CRediT\footnote{\url{https://www.elsevier.com/authors/policies- and- guidelines/credit- author- statment}}).\\
\textbf{Matthew~A.~Price:}
\noindent Conceptualisation, Methodology, Software, Investigation, Validation, Writing (Original Draft, Review \& Editing, Supervision)
\textbf{Jason~D.~McEwen:}
\noindent Conceptualisation, Methodology, Writing (Review \& Editing, Supervision).
\textbf{Alicja~Polanska \& Jessica~Whitney:}
\noindent Software, Investigation, Validation.

\bibliographystyle{mymnras_eprint}
\bibliography{bib,mybibs_new}

\clearpage
\appendix
\section{Wavelet Construction} \label{sec:wav_construct}
Scale-discretised wavelets on the sphere and ball ${}_s\Psi^{jj^\prime}$ provide a natural dictionary in which many physical fields are sparsely distributed. For exact decomposition and synthesis into and from a wavelet dictionary respectively ${}_s\Psi^{jj^\prime}$ must satisfy the admissibility condition
\begin{equation} \label{eq:admissibility}
  \frac{4\pi}{2\ell+1}\:|{}_s\phi_{\ell0,p}|^2
  + \frac{8\pi^2}{2\ell+1}\sum_{ijm}|{}_s\Psi^{jj^\prime}_{\ell m, p}|^2 = 1,
\end{equation}
for all $\ell$ and $p$. When considering functions with no radial extent one may simply ignore both $p$ and radial scales $j^\prime$. A key factor which increases the efficacy of wavelets is their localisation properties, and so careful consideration of how ${}_s\Psi^{jj^\prime}$ are constructed is critically important. In this section we consider the scale-discretised wavelets which are separable in the harmonic $\ell$, azimuthal $m$, and radial $p$ components and satisfy the admissibility condition in Equation \ref{eq:admissibility}. In the following sections we discuss one such choice of these tiling functions which has proved to be particularly effective.

\subsection{Tiling of the harmonic line} \label{sec:tiling_harmonic}
To capture information along the harmonic line $\ell$ we define positive real functions $\kappa^{(j)}(t)$ for $t \in \mathbb{N}^+$ such that $t \leq L$, where $j$ denotes angular wavelet scale. The construction of this function is given by the following. First, consider the infinitely differentiable $C^{\infty}$ Schwartz function
\begin{equation}
  s_{\lambda}(t)
  = s \:
  \Big (
  \frac{2\lambda}{\lambda -1}(t-\frac{1}{\lambda})-1
  \Big),
\end{equation}
with compact support $t\in[\lambda^{-1},1]$ and for dilation parameter $\lambda \in \mathbb{R}^+_{>1}$. Note that the most common choice of dilation parameter is $\lambda = 2$ in which case we recover dyadic wavelets. The function $s(t)$ is given by
\begin{equation}
  s(t)
  = \exp\big(-\frac{1}{1-t^2} \big ),
\end{equation}
for $t\in[-1,1]$ and 0 elsewhere. From these functions with compact support we can define the function
\begin{equation}
  k_{\lambda}(t)
  = \frac{
  \int_t^1 \frac{\text{d}t^\prime}{t^\prime}s_{\lambda}^2(t^\prime)
  }{
  \int_{\lambda^{-1}}^1 \frac{\text{d}t^\prime}{t^\prime} s_{\lambda}^2(t^\prime),
  }
\end{equation}
which decays smoothly from unity at $t < \lambda^{-1}$ to zero at $t > 1$. We can now focus the localisation in $t$ by defining
\begin{equation}
  \kappa_{\lambda}(t) = \sqrt{k_{\lambda}(\lambda^{-1}t) - k_{\lambda}(t)},
\end{equation}
which is the generating function for the harmonic component of the wavelet construction, with which the scale-discretised wavelet harmonic kernel for scale $j$ is defined by $\kappa^{(j)}(\ell) = \kappa_{\lambda}(\lambda^j L^{-1}\ell)$,
which is compact on $\ell \in \big [ \lfloor\lambda^{-(1+j)}L\rfloor, \lceil\lambda^{1-j}L\rceil \big ]$. With this construction, it is clear that each wavelet scale need only be evaluated between some lower and upper harmonic degrees, which we will denote hereon out as $L_{j^\pm}$ respectively. To capture low frequency information one must also introduce a scaling tiling function $\eta_{\lambda}(t) = \sqrt{k_{\lambda}(t)}$.

\subsection{Tiling of the azimuthal line} \label{sec:tiling_azimuthal}
To capture the azimuthal (directional) information content of functions on the sphere we define the spin-$s$ square integrable function ${}_s\zeta \in \text{L}^2[\mathbb{S}^2]$ with harmonic representation $\langle {}_s\zeta, {}_sY_{\ell m} \rangle$, following Equation \ref{eq:fswsh}. An azimuthal bandlimited $N$ is introduced such that ${}_s\zeta_{\ell m} = 0$ for all $\ell, m$ where $|\,m\,| \geq N$, in much the same way as was done for bandlimited functions on the sphere. The azimuthal tiling function is defined as
\begin{equation}
  {}_s\zeta_{\ell m} = \xi\mu \sqrt{\frac{1}{2^q} \binom{q}{(q-m)/2}},
\end{equation}
which is a specific form of the directional auto-correlation with $\xi=1$ for even $N-1$ and $\xi=i$ else, and $\mu = [1-(-1)^{N+m}]/2$ and $q = \min \lbrace N-1, \ell-[1+(-1)^{N+\ell}]/2\rbrace$. The full details of this derivation are extremely involved and can be found in related work.

\subsection{Tiling of the ball} \label{sec:tiling_ball}
The wavelet generating function on the radial half-line is in fact identical to that of the harmonic line, the only difference being that here we consider tiling over the associated $2^{\text{nd}}$-order Laguerre polynomials indexed by $p$. Therefore identical logic to that of Section \ref{sec:tiling_harmonic} may be applied to derive the radial tiling function $\kappa^{(j^\prime)}(p)$ and scaling function $\eta(p)$, where we define a radial dilation parameter $\lambda \rightarrow \nu$.

Combining this radial tiling and the harmonic tiling discussed in Sections \ref{sec:tiling_harmonic} one may construct a hybrid generating function for scale-discretised wavelets on the ball, given simply by
\begin{equation}
  \kappa^{(jj^{\prime})}(\ell, p) = \kappa^{(j)}(\ell) \: \kappa^{(j^\prime)}(p).
\end{equation}
We finally define the hybrid scaling function generator
\begin{equation}
  \eta_{\lambda \nu}(t, t^\prime)
  = \sqrt{k_\lambda\Big(\frac{t}{\lambda}\Big)k_{\nu}(t^\prime) + k_{\lambda}(t)k_{\nu}\Big(\frac{t^\prime}{\nu}\Big) - k_{\lambda}(t)k_{\nu}(t^\prime)},
\end{equation}
which both captures low frequency information and ensures the constructed dictionary on the ball satisfies the wavelet admissibility condition.

\subsection{Scale-discretised Wavelet Dictionary}
Combining the results from this section we can explicitly define our directional spin-$s$ wavelet functions in their harmonic representation
\begin{equation}
  {}_s \Psi^{jj^\prime}_{\ell m p}
  = \sqrt{\frac{2\ell+1}{8\pi^2}} \:
  \kappa^{j}(\ell) \:
  \kappa^{j^\prime}(p) \:
  {}_s\zeta_{\ell m}.
\end{equation}
The corresponding scaling functions ${}_s\Phi$ are given by
\begin{equation}
  {}_s\Phi_{\ell 0 p}
  = \begin{cases}
    \sqrt{\frac{2\ell+1}{4\pi}}\eta_{\nu}(\frac{p}{\nu^{J_0}})                                           & \text{if $\ell > \lambda^{J^\prime_0}, p \leq \nu^{J_0}$} \\
    \sqrt{\frac{2\ell+1}{4\pi}}\eta_{\lambda}(\frac{\ell}{\lambda^{J^\prime_0}})                         & \text{if $\ell \leq \lambda^{J^\prime_0}, p > \nu^{J_0}$} \\
    \sqrt{\frac{2\ell+1}{4\pi}}\eta_{\lambda\nu}(\frac{\ell}{\lambda^{J^\prime_0}}, \frac{p}{\nu^{J_0}}) & \text{if $\ell < \lambda^{J^\prime_0}, p < \nu^{J_0}$},
  \end{cases}
\end{equation}
and zero elsewhere. Here $J_0$ and $J^\prime_0$ denote the minimum wavelet scales $jj^\prime$, below which the scaling function captures any remaining information.

\end{document}